\documentclass[article,ij4uq]{ij4uq}              %  final version
\DeclareSymbolFont{matha}{OML}{txmi}{m}{it}% txfonts
\DeclareMathSymbol{\varv}{\mathord}{matha}{118}
\usepackage{color}
\usepackage{subcaption}
\usepackage{xspace}
\usepackage{xifthen}
\usepackage{multicol}
\usepackage{mathtools}
\usepackage{bbm}
\usepackage{wasysym}%smileys
\usepackage[makeroom]{cancel}
\usepackage[english]{babel}
\usepackage{graphicx}
\usepackage{fontenc}
\usepackage{amssymb}
\usepackage{dsfont}
\usepackage{placeins}
\usepackage{xcolor}
\usepackage{enumerate}
\usepackage{tabularx}
\usepackage{csquotes}
\usepackage{ mathrsfs }

\usepackage{comment}
\usepackage{nomencl}
\usepackage{makecell}
\usepackage{geometry}
\usepackage{booktabs}
\usepackage{cleveref}
\usepackage{tikz}
\usepackage{pgfplots}
\usepackage{mathrsfs}
\usepackage[hang]{footmisc}
\fancypagestyle{plain}{%
  \fancyhf{}
  \fancyhead[R]{\small {\it International Journal for Uncertainty Quantification}, x(x): \thepage--\pageref{LastPage} (\myyear\today)}
  \fancyfoot[R]{\small\bf\thepage }
  \fancyfoot[L]{\fottitle}
  }
\renewcommand{\dmyy}{17}
\renewcommand{\myyear}{2017}
\renewcommand{\today}{}

\newcommand{\x}{\mathbf{x}}
\newcommand{\lambd}{\boldsymbol{\lambda}}

\newcommand{\Lambd}{\mathbf{\Lambda}}
\newcommand{\Lambdprime}{\Lambd'}
\newcommand{\alpp}{\boldsymbol{\alpha}}

\newcommand{\f}{\boldsymbol{f}}

\newcommand{\yobsj}{y_{1,\text{obs},j}}
\newcommand{\Yobsj}{Y_{1,\text{obs},j}}
\newcommand{\Yobskj}{Y_{t',\text{obs},j}}
\newcommand{\yobskj}{y_{t',\text{obs},j}}
\newcommand{\yobs}{\boldsymbol{y_{1,\text{obs}}}}
\newcommand{\yobsloo}{\boldsymbol{y_{1,\text{obs}}^{-j}}}
\newcommand{\Yobs}{\boldsymbol{Y_{1,\text{obs}}}}
\newcommand{\Yobsk}{\boldsymbol{Y}_{\boldsymbol{t'},\textbf{obs}}}
\newcommand{\yobsk}{\boldsymbol{y}_{\boldsymbol{t'},\textbf{obs}}}

\newcommand{\eps}{\boldsymbol{E}}
\newcommand{\epsj}{E_j}
\newcommand{\epsf}{E^f}
\newcommand{\epstot}{E^{\mathrm{tot}}(\lambd)}
\newcommand{\Abf}{\mathbf{A}}
\newcommand{\alpmap}{\alpp_{\text{MAP}}}

\newcommand{\Xii}{\boldsymbol{\Xi}}
\newcommand{\xii}{\boldsymbol{\xi}}
\newcommand{\HH}{\mathbf{H}}

\newcommand{\FF}{\hat{\mathbf{F}}}

\newcommand{\y}{\boldsymbol{y}}
\newcommand{\Y}{\boldsymbol{Y}}
\newcommand{\discrep}{\Delta}
\newcommand{\Lcal}{\mathcal{L}}
\newcommand{\Lphy}{\mathcal{L}_{\text{cal}}}
\newcommand{\Lnum}{\mathcal{L}_{\text{err}}}
\newcommand{\normx}[1]{\lVert #1 \lVert}
\newcommand{\ntrain}{n_{\mathrm{train}}}

\newcommand{\perturb}{\omega}
\newcommand{\lambdzero}{\lambd_{0}}
\newcommand{\ind}{\perp\!\!\!\!\perp} 
\newcommand{\thresh}{\tau}
\newcommand{\dist}{\nu}
\newcommand{\alppstarl}{\alpp^{\star}_{\ell}}
\newcommand{\alppstarlj}{\alpp^{j\star}_{\ell}}
\newcommand{\ENM}[1]{\hat{E}_{N,M}^{\alppstarl}\Big(#1\Big)}

\newcommand{\ENMj}[1]{\hat{E}_{N,M}^{\alppstarlj}\Big(#1\Big)}

\newcommand{\EM}[1]{\hat{E}_M\Big(#1\Big)}
\newcommand{\alpinter}{\kappa}

\newcommand{\Ma}{L}
\newcommand{\Kset}{\mathcal{K}}

\begin{document}

\volume{Volume x, Issue x, \myyear\today}
\title{Bayesian Calibration for prediction in a multi-output transposition context}
\titlehead{Bayesian Calibration in a multi-output transposition context}
\authorhead{C. Sire, J. Garnier, B. Kerleguer, C. Durantin, G. Defaux, G. Perrin}
%For at least  authors with different addresses, use instead the following commands
\corrauthor[1,2]{Charlie Sire}
\author[2]{Josselin Garnier}
\author[3]{Baptiste Kerleguer}
\author[3]{Cédric Durantin}
\author[3]{Gilles Defaux}
\author[4]{Guillaume Perrin}
\corremail{sire.charlie971@gmail.com}
\corraddress{Inria Saclay Centre, Palaiseau, France}
\address[1]{Inria Saclay Centre, Palaiseau, France}
\address[2]{CMAP, CNRS, Ecole polytechnique, Institut Polytechnique de Paris, 91120 Palaiseau, France}
\address[3]{CEA, DAM, DIF, F-91297 Arpajon, France}
\address[4]{COSYS, Universite Gustave Eiffel, Marne-La-Vallée, France}
% End information for at least  authors with different addresses
% For authors with the same post address,
%\corrauthor{First A. Author}
%\corremail{f.author@affiliation.com}
%\author{Second B. Author, Jr.}
%\address{Department of Chemistry and Courant, Institute of Mathematical Sciences, New York, NY 10012, USA}
% End commands for all authors with the same address

\dataO{mm/dd/yyyy}
%\dataO{}
\dataF{mm/dd/yyyy}
%\dataF{}

\abstract{
Numerical simulations are widely used to predict the behavior of physical systems, with Bayesian approaches being particularly well suited for this purpose. 
However, experimental observations are necessary to calibrate certain simulator parameters for the prediction. 
In this work, we use a multi-output simulator to predict all its outputs, including those that have never been experimentally observed. 
This situation is referred to as the transposition context. 
To accurately quantify the discrepancy between model outputs and real data in this context, conventional methods cannot be applied, and the Bayesian calibration must be augmented by incorporating a joint model error across all outputs. 
To achieve this, the proposed method is to consider additional input parameters within a hierarchical Bayesian model, which includes hyperparameters for the prior distribution of the calibration variables. 
This approach is applied to a computer code with three outputs that models the Taylor cylinder impact test with a small number of observations. 
The outputs are considered as the observed variables one at a time, to work with three different transposition situations. 
The proposed method is compared with other approaches that embed model errors to demonstrate the significance of the hierarchical formulation.}


\maketitle

\section{Introduction}
\label{intro}

Numerical simulations are increasingly used to understand and optimize complex physical systems, as they enable the prediction of behavior in previously unobserved scenarios.
For accurate predictions, the input parameters of these simulations must be carefully specified.
Some of the simulation inputs, known as control variables, replicate experimental conditions chosen by the experimenter and are denoted by $\x$, while others are unknown parameters specific to the computer code, denoted by $\lambd$.
These parameters typically have a physical significance and must be calibrated to ensure the code predictions closely match the actual system behavior, observed from limited experimental data. 
The parameters $\lambd$ can also be numerical parameters that are needed to run the code. The simplest approach involves performing an optimization scheme on $\lambd$ to minimize a given distance between observed data and model outputs, see \cite{garrett,reddy,muehleisen}, then using the optimal values as if they were known. However, this method assumes no uncertainty in the observed data (no measurement error) and in the code prediction (no model error), which is unrealistic in practice.

A Bayesian methodology that considers noisy observations and treats the unknown parameters as a random vector $\Lambd$ can be adopted. The prior distribution of $\Lambd$ and the observations are used to build the posterior distribution of $\Lambd$, which captures the uncertainty and allows for a predictive distribution of the output.
This framework is thoroughly presented in \cite{kennedy} and further studied in numerous works, for example~\cite{higdon,bayarri,boukouvalas}.

Nevertheless, predicting the behavior of physical systems using numerical models is never perfect, even with well-identified parameter values. There is always a discrepancy between the model outputs and the observed data, which must be accounted for as a model error, as highlighted in various works, see \cite{CAMPBELL20061358,OHagan2013,ling2014selection,maupin2020model}. 
Typically, as detailed in \cite{kennedy}, the model error term can be explicitly represented as an additive Gaussian process, whose distribution hyperparameters are estimated from the observed data. This approach is particularly useful but cannot be implemented for prediction in some situations due to the impossible estimation of the hyperparameters. These situations arise when the output of interest does not directly correspond to the experimental data, such as when calibration must be performed based on different experiments not directly suited to the objective.
The framework introduced in this paper will be called transposition. Four cases can be identified:
\begin{enumerate}
  \item Change of scale: The calibration domain is different from the domain of interest of the code.
  \item New inputs: The code is calibrated with fewer inputs than it will use.
  \item New code: The calibration parameters will be used in another code.
  \item Unobserved outputs: The calibration and prediction will be carried out by observing some outputs, but not all of them.
\end{enumerate}
In this paper, only the case of predicting unobserved outputs will be considered, although the presented method is general enough to be adapted to many transposition scenarios.

A collection of $T$ different output variables of interest is identified and denoted by $\y$. It depends on control variables $\x \in \mathcal{X}$, with $\mathcal{X}$ a subset of $\mathbb{R}^s, s \in \mathbb{N}^\star,$ such that:
$$\y(\x) = (y_t(\x))_{t=1}^T \in \mathbb{R}^T.$$
Let $\f$ be a numerical code to compute this $T$-dimensional output. Its input variables are  $\x $ and $p$ unknown physical calibration parameters $\lambd = (\lambda_i)_{i=1}^p \in \Lcal \subset \mathbb{R}^p$, and
$$\f(\x, \lambd) = (f_t(\x, \lambd))^T_{t=1} \in \mathbb{R}^T.$$
The experimental phase is assumed to provide $n$ observations but only for the variable $t'$, where $1\leq t'\leq T.$ They form the vector $\yobsk = (\yobskj)_{j=1}^n$ which we consider as realizations  of noisy measurements of $y_{t'}$ at design points $\mathbb{X} = (\x_j)_{j=1}^n$ with output measurement error $\eps = (\epsj)_{j=1}^n$:

\begin{equation}
\Yobsk = (\Yobskj)_{j=1}^n = (y_{t'}(\x_j) + \epsj)_{j=1}^n \in \mathbb{R}^n. 
\end{equation}

The probability density function (pdf) of the measurement error is known and is denoted by $p_{\eps}$. 
The transposition comes from the fact that only the variable $y_{t'}$ is measured in the experimental phase, while the objective is to predict $y_t(\x_0)$ for $1 \leq t \leq T$, and $\x_0 \in \mathcal{X}\setminus \mathbb{X}$. In the following, and without loss of generality, we assume that the variable $y_1$ is measured (i.e. $t' = 1$) considering the random vector $\Yobs = (\Yobsj)_{j=1}^n = (y_{1}(\x_j) + \epsj)_{j=1}^n.$ Note that in the proposed framework, the number of experiments $n$ is assumed to be small (e.g., $n=10$ in the application presented in Section~\ref{appli}). This limited sample size makes prediction at $\x_0$ more challenging. However, this constraint is often unavoidable due to the financial, material, or time-related costs associated with each physical experiment.

The common additive discrepancy to represent the model error would be introduced as follows, by modeling the output variables of interest as
$\Y(\x) = (Y_t(\x))_{t=1}^T = \f(\x, \Lambd) + \boldsymbol{\discrep}(\x)$, with $\Lambd \in \mathbb{R}^p$ a random vector of pdf $p_\Lambd$ and $\boldsymbol{\discrep}$ a $T$-variate Gaussian process with prior information depending on some hyperparameters $\phi$. In our transposition context, some of these hyperparameters are impossible to estimate. Indeed, let us consider for instance $\boldsymbol{\discrep} = (\discrep_t)_{t=1}^T$ pairwise independent Gaussian processes with respective mean $\mu_t(.)$ and covariance $c_t(.,.)$ that depends on hyperparameters $\phi_t$. It quickly appears that for every $t$, experimental measurments of $y_t$ are needed to estimate $\phi_t$ , but they are not available in this context.

Therefore, the objective is to propose a model error for the prediction of $y_t(\x_0)$ with hyperparameters that can be estimated from the observations $\Yobs$. As presented in \cite{OLIVER20151310}, embedded errors can be employed to build a mathematical relationship between the observations and the variables of interest. A similar approach is described in \cite{huan_embedded}, where an embedding $Y_t(\x) = f_t(\x, \Lambd + \boldsymbol{\Delta})$ is considered with $\boldsymbol{\Delta}$ being a random discrepancy whose parameters require calibration, as detailed in \ref{embed}. In our work, an original embedding is proposed by increasing the variability in the calibration variables $\Lambd$ through the inclusion of $q-p$ additional parameters to represent the model error, and then work with a $q$-dimensional vector $\Lambd$. These $q-p$ parameters encompass all parameters that influence the relationship between the controlled inputs and all the $T$ outputs but are not classically intended to be calibrated. These could include, for example, mesh parameters, variables related to the simulation duration, or known physical law parameters.  
As this study is conducted in the context of a very limited number of observations, prior assumptions play a crucial role. This is why a hierarchical approach is proposed for these additional parameters, allowing for the estimation of the hyperparameters of their prior distribution \cite{DAMBLIN2020110391}, thereby refining it to better account for model error. 

The main contribution of this work lies in this embedded model error for prediction in transposition contexts. A key challenge is the treatment of hyperparameters within the hierarchical framework.
Note that this method cannot be effective if the variations in the output due to the fluctuations of the additional parameters are not sufficiently large to capture the model error. This point will be further discussed in \Cref{appli} and \ref{study_prior}.

The article is organized as follows: \Cref{sec_bayes} outlines the Bayesian methodology tailored to the transposition context and introduces the surrogate model used. \Cref{appli} describes the application test case and the performance metrics examined, followed by numerical results with a comparison of various approaches, including the method from \cite{huan_embedded}. Finally, \Cref{conclu} summarizes the work and proposes possible extensions.

\section{Building an adapted Bayesian framework}\label{sec_bayes}

\subsection{Bayesian calibration}\label{bayes_simple}

In the Bayesian framework, the idea is to obtain the posterior distribution (i.e. conditioned by the experimental data) of parameters, that balances prior knowledge with observed data, see \cite{van2021bayesian}. More precisely, we consider
\begin{equation}\label{eq_bay_err}
\left\{
\begin{aligned}
    Y_t(\x) &= f_t(\x, \Lambd), ~ \: ~ \: ~ \: ~ \: ~ \: \:\: 1\leq t\leq T, ~ \x \in \mathcal{X}\\
    \Yobsj &= Y_1(\x_j) + \epsj, ~ \: ~ \: \: 1\leq j \leq n,
\end{aligned}
\right.
\end{equation}
with $\Lambd$ of prior pdf $p_\Lambd$, $\eps$ of known pdf $p_{\eps}$, and a small $n.$
As explained in \Cref{intro}, here we consider $\Lambd \in \mathbb{R}^q$, that is made of the $p$ unknown physical parameters and the $q-p$ additional parameters included to emulate the model error. Obviously, the following formulas are valid when considering the classical situation $\Lambd \in \mathbb{R}^p$.

The goal is to predict $y_t(\x_0)$, which requires investigating the posterior distribution of $Y_t(\x_0)$, i.e., the distribution of $Y_t(\x_0)\mid \yobs$. This involves estimating $\mathbb{E}(\psi(Y_t(\x_0)) \mid \yobs)$ for all functions $\psi: \mathbb{R} \to \mathbb{R}$ such that $\mathbb{E}(\psi(Y_t(\x_0)) \mid \yobs)$ is well-defined. In particular, the estimation of $\mathbb{E}(Y_t(\x_0) \mid \yobs)$ and $\mathbb{E}(Y_t(\x_0)^2 \mid \yobs)$ are required to determine the predictive mean and standard deviation. In situations where extreme events are studied, functions of the form $\Psi(y) : y \to \mathbf{1}_{[b, +\infty[}$, with $b\in \mathbb{R}$, are considered.

The expectation given the observations can be expressed:
\begin{equation}\label{eq_post}
\mathbb{E}(\psi(Y_t(\x_0)) \mid \yobs) = \int_{\mathbb{R}^q} \psi(f_t(\x_0, \lambd)) p(\lambd \mid \yobs) d\lambd,
\end{equation}
with $p(\lambd \mid \yobs)$ the pdf of the posterior distribution of $\Lambd$. 
Bayes' theorem, see \cite{berkson1930bayes}, gives $p(\lambd \mid \yobs) = \frac{p(\yobs \mid \lambd)p_{\Lambd}(\lambd) }{\int_{\mathbb{R}^q} p(\yobs \mid \lambd')p_{\Lambd}(\lambd') d\lambd'}$, where $p(\yobs \mid \lambd) = p_{\eps}\left(\yobs - (f_1(\x_j,\lambd))_{j=1}^n \right)$ is the likelihood of $\lambd$. A consistent estimator of $\mathbb{E}(\psi(Y_t(\x_0) \mid \yobs)$ can be
\begin{equation}\label{estim_M}
\EM{\psi(Y_t(\x_0))} = \frac{1}{M}\sum_{k=1}^{M}\psi(f_t(\x_0, \Lambd_k)),
\end{equation}
where $(\Lambd_k)_{k=1}^M$ is sampled with pdf
$p(\lambd \mid \yobs)\propto p(\yobs \mid \lambd)p_{\Lambd}(\lambd)$. However, the normalizing term $\int_{\mathbb{R}^q} p(\yobs \mid \lambd')p_{\Lambd}(\lambd') d\lambd'$ is known to be intractable in practice, making it impossible to sample directly from $p(\lambd \mid \yobs)$ with classical Monte Carlo schemes. Markov Chain Monte Carlo (MCMC) methods are then required here to work with posterior samples $(\Lambd_k)_{k=1}^{M}$ with pdf proportional to $p(\yobs \mid \cdot )p_{\Lambd}.$ In our study, these MCMC samples are obtained with the Delayed Rejection Adaptive Metropolis (DRAM) algorithm detailed in \cite{haario2006dram}, but other algorithms such as Hamiltonian Monte Carlo (HMC, \cite{DUANE1987216}) could be used, in particular if $q$ is large. 

Note that here $p_{\eps}$ will be considered known in all the methods we implement, to ensure a fair comparison between them and focus on the model error. However, in some situations, inferring the parameters of $p_{\eps}$ is crucial. For instance, if $\eps \sim \mathcal{N}(0, \sigma_\varepsilon^2I)$ with unknown $\sigma_\varepsilon$, then a solution is to consider $\sigma_\varepsilon$ as a hyperparameter to be inferred together with $\lambd$, and then work with the augmented set $\tilde{\lambd} = (\lambd, \sigma_\varepsilon)$ of likelihood $p(\tilde{\lambd}\mid \yobs) = p_{\eps(\sigma_\varepsilon)}\left(\yobs - (f_1(\x_j,\lambd))_{j=1}^n \right)$ with $p_{\eps(\sigma_\varepsilon)}$ the density of $\eps(\sigma_\varepsilon) \sim \mathcal{N}(0, \sigma_\varepsilon^2I).$

\subsection{Hierarchical model}\label{hier}

In the following, $\Lphy \subset \mathbb{R}^p$ is the domain of physical parameters to calibrate, and $\Lnum \subset \mathbb{R}^{q-p}$ is the domain of the additional parameters considered to emulate the model error.
Then $\lambd \in \Lcal$ is introduced with $\Lcal = \Lphy \times \Lnum$. 
In our study, the $q-p$ parameters in $\Lnum$ are integrated with a hierarchical description, see \cite{kemp2007learning} and \cite{allenby2006hierarchical}, to better target their correct distribution and account for the model error. This is particularly effective even in contexts where measurement noise is significant. The hierarchical description assumes that the a priori distribution of the additional parameters depends on hyperparameters $\alpp \in \mathcal{A} \subset \mathbb{R}^r$.
The pdf of $\Lambd$ given $\alpp$ is denoted $p_{\Lambd}(. \mid \alpp),$ and each $\alpp$ drives these parameters for the model error. In a hierarchical description, the objective is to perform Bayesian inference on the hyperparameters $\alpp$ as well, and then consider them as a random variable $\Abf \in \mathcal{A}$ with prior distribution $p_{\Abf}$. Note that, to avoid possible confusion, $\Abf$ is simply the upper case notation of $\alpp$ here, as lower-case letters represent deterministic variables, while the upper-case letters are associated with random variables throughout  this article. The same convention applies to $\lambd$ and $\Lambd$ for instance. 

As for \Cref{eq_post}, Bayes' Theorem provides that $p(\alpp \mid \yobs)$, the posterior pdf of $\Abf$ evaluated at $\alpp$, is proportional to $p(\yobs \mid \alpp)p_{\Abf}(\alpp)$, where $p(\yobs \mid \alpp)$ is the likelihood of $\alpp$. This likelihood can be expressed as 
\begin{equation}
\begin{aligned}
    p(\yobs \mid \alpp) &= \int_{\mathbb{R}^q} p(\yobs \mid \lambd,\alpp)p_{\Lambd}(\lambd \mid \alpp)d\lambd \\ &= \int_{\mathbb{R}^q} p(\yobs \mid \lambd)p_{\Lambd}(\lambd \mid \alpp)d\lambd ~~\text{ as given }\lambd, \yobs\text{ does not depend on }\alpha\\
    &= \mathbb{E}\left[p(\yobs \mid \Lambd)\mid \alpp\right]
\end{aligned}
\end{equation}

\noindent which is the expectation of $p(\yobs \mid \Lambd)$ when $\Lambd$ has pdf $p_{\Lambd} (.\mid \alpp).$
Thus, the most natural way to estimate $p(\yobs \mid \alpp)$ for a given $\alpp$ is to compute the Monte Carlo estimator 
\begin{equation}
\hat{P}_{\Ma}(\yobs \mid \alpp) = \frac{1}{\Ma}\sum_{k=1}^{\Ma} p(\yobs \mid \Lambdprime_k)
\end{equation}

\noindent with $(\Lambdprime_k)_{k=1}^{\Ma}$ i.i.d. with pdf $p_{\Lambd}(. \mid \alpp).$
With this approach, it would lead to $\Ma \times n$ runs of the computer code for the estimation of the likelihood of a single $\alpp$, which is not feasible in practice.
To save significant computation time, the idea here is to use an Importance Sampling formulation, presented in \cite{kloek}, based on the following result:
\begin{flalign*}
&\forall (\alpp, \alpp^{\star}) \text{ such that } \text{supp}(p_{\Lambd}(. \mid \alpp)) \subset \text{supp}(p_{\Lambd}(. \mid \alpp^\star)), &&\\
&\mathbb{E}\left[p(\yobs \mid \Lambd) \mid \alpp\right] =\mathbb{E}\left[p(\yobs \mid \Lambd) \frac{p_{\Lambd}(\Lambd \mid \alpp)}{p_{\Lambd}(\Lambd \mid \alpp^\star)} \mid \alpp^\star\right]
\end{flalign*}

Then, for all $\alpp$, we can use the following Importance Sampling estimator of $p(\yobs \mid \alpp)$, based on sampling associated with a vector of hyperparameters $\alpp^\star$:
\begin{equation}\label{estimator_likeli}
\hat{P}^{\alpp^\star}_{\Ma}(\yobs \mid \alpp) = \frac{1}{\Ma}\sum_{k=1}^{\Ma} p(\yobs \mid \Lambdprime_k) \frac{p_{\Lambd}(\Lambdprime_k \mid \alpp)}{p_{\Lambd}(\Lambdprime_k \mid \alpp^\star)},
\end{equation}

\noindent with $(\Lambdprime_k)_{k=1}^\Ma$ i.i.d. with pdf $p_{\Lambd}(. \mid \alpp^\star).$ This estimator is consistent and asymptotically normal, as detailed in \Cref{alpmap}.
Here, the same samples are used for all $\alpp$. Then, with only $L \times n$ runs of the computer code we can estimate the likelihood $p(\yobs \mid \alpp)$ for all $\alpp$, rather than just for a single $\alpp$ with the Monte Carlo estimation. This allows us to conduct a plug-in strategy by maximizing the posterior density of $\Abf$ and working with the maximum a posteriori $\alpmap  = \underset{\alpp \in \mathcal{A}}{\text{argmax}}\: p(\alpp\mid \yobs)$. Another strategy is to adopt a full-Bayesian approach and consider the entire posterior distribution of $\Abf$ with MCMC methods.

\subsection{Investigating $\alpmap$}\label{alpmap}

This section details the estimation of $\alpmap$ which can then be used as a plug-in value for the hyperparameters to compute $\EM{\psi(Y_t(\x_0))}$ with the prior distribution $p_{\Lambd}(.\mid \alpmap)$ for $\Lambd.$ The idea is to consider $\hat{P}^{\alpp^\star}_{\Ma}(\yobs \mid \alpp)$ as the estimator of the likelihood $p(\yobs\mid \alpp)$ of $\alpp,$ for a relevant $\alpp^\star$ as detailed in the following. 

\subsubsection{Estimation of $\alpmap$.}

Theoretically, the maximization of the posterior density on $\mathcal{A}$ can be performed by selecting a given $\alpp^{\star}$ and evaluating $\hat{P}^{\alpp^\star}_{\Ma}(\yobs\mid\alpp)p_{\Abf}(\alpp)$ for every $\alpp$. However, the ratios $\frac{p_{ \Lambd}(\Lambdprime_k\mid \alpp)}{p_{ \Lambd}(\Lambdprime_k\mid \alpp^\star)}$ can be highly fluctuating when $\alpp$ is far from $\alpp^\star$, leading to estimators with a high variance. To overcome this issue, the proposed solution is work with $\alpp^\star$ close to $\alpmap$, using an iterative strategy in $\alpp^\star$ until reaching a fixed point, as detailed in \Cref{alg-map}.
The inputs of the algorithm are $\Abf^{\star}_0$, the initial value of $\alpp^{\star}$ and $\thresh$, the stop threshold. The value of $\thresh$ depends on the interpretation of $\alpp$, and will be briefly discussed in Section~\ref{methods}.
Note that in this work, the optimization $\Abf^{\star}_{\ell+1} = \underset{\alpp \in \mathcal{A}}{\text{argmax}}\:\hat{P}^{\Abf^{\star}_{\ell}}_{\Ma}(\yobs\mid\alpp)p_{\Abf}(\alpp)$ is performed with L-BFGS-B optimizer \cite{liu1989limited}. In the following, we will work with the optimal vector $\alppstarl,$ which is the approximation of $\alpmap$ obtained at the end of the algorithm. 

\begin{algorithm}[!t]
    \label{}
    \textbf{Input:} $\Abf^{\star}_0$, $\thresh$\\
        $\ell \gets 0$, $\dist \gets +\infty$\\
        \While{$ \dist > \thresh$}{
            Sample $(\Lambdprime_k)_{k=1}^{\Ma}$ i.i.d. with  pdf $p_{\Lambd}(.\mid \Abf^{\star}_{\ell})$\\
            $\Abf^{\star}_{\ell+1} \gets \underset{\alpp \in \mathcal{A}}{\text{argmax}}\:\hat{P}^{\Abf^{\star}_{\ell}}_{\Ma}(\yobs\mid\alpp)p_{\Abf}(\alpp)$\\
            $\dist \gets \normx{\Abf^{\star}_{\ell} - \Abf^{\star}_{\ell+1}}$\\
            $\ell \gets \ell +1$
       }
\caption{Iterative estimation of $\alpmap$}\label{alg-map}
\end{algorithm}

\subsubsection{Confidence in the estimation.}\label{conf_in_est}

Although this is not required in the proposed method, it is possible to check that $p(\alppstarl \mid \yobs)$, the posterior pdf at $\alppstarl,$ is close to the maximum a posteriori density across all $\alpp \in \mathcal{A}$. 
As detailed in \ref{clt}, with a statistical test, we propose a confidence level $\gamma(\alpp)$ associated with $p(\alpp \mid \yobs) < \beta p(\alppstarl \mid \yobs)$ for a given $\alpp$, where $\beta$ represents an acceptable margin of error for this estimation, set by the user. For example, in the application, $\beta = 1.05$ is used, see \Cref{numerical_results}, corresponding to a tolerance of $5\%$ on the posterior density. This confidence level will be computed for $\alpp \in \mathcal{A}$, as illustrated in \Cref{confidence_alpha}. A threshold $\zeta$ can be introduced, to ensure that $\forall \alpp \in \mathcal{A}, \gamma(\alpp) \geq \zeta,$ with for instance $\zeta = 0.95.$ To estimate these confidence levels and similarly to the approach detailed in \Cref{full_bayes}, we use $\hat{P}^{\alppstarl}_{\Ma}(\yobs\mid\alpp)$ as the estimator of $p(\yobs\mid\alpp).$ More precisely, for this specific study, a number $\Ma'$ of samples is considered, with $\Ma'$ very large, to work with $\hat{P}^{\alppstarl}_{\Ma'}(\yobs\mid\alpp)$ and investigate the convergence, when $\Ma'$ goes to $+\infty$, of 
\begin{equation}\label{eq_tcl}
\begin{aligned}
\hat{P}^{\alppstarl}_{\Ma'}(\yobs\mid\alpp)p_{\Abf}(\alpp) 
&- \beta\hat{P}^{\alppstarl}_{\Ma'}(\yobs\mid\alppstarl)p_{\Abf}(\alppstarl) \\
&= \frac{1}{\Ma'}\sum_{k=1}^{\Ma'} \left( p(\yobs \mid \Lambdprime_k)
\frac{p_{\Lambd}(\Lambdprime_k\mid \alpp)p_{\Abf}(\alpp) 
- \beta p_{\Lambd}(\Lambdprime_k\mid \alppstarl)p_{\Abf}(\alppstarl)}{p_{\Lambd}(\Lambdprime_k\mid \alppstarl)}\right),
\end{aligned}
\end{equation}

with $(\Lambdprime_k)_{k=1}^{\Ma'}$ i.i.d. with pdf $p_{\Lambd}(. \mid \alppstarl).$

We introduce the estimator
\begin{align*}
S_{L'}(\alpp)^2 = &\frac{1}{L'-1}\sum_{k=1}^{L'} \Bigg( p(\yobs \mid \Lambd'_k)\frac{p_{\Lambd}(\Lambd'_k\mid \alpp)p_{\Abf}(\alpp) - \beta p_{\Lambd}(\Lambd'_k\mid \alppstarl)p_{\Abf}(\alppstarl)}{p_{\Lambd}(\Lambd'_k\mid \alppstarl)}- \\
&\left(\hat{P}^{\alppstarl}_{\Ma'}(\yobs\mid\alpp)p_{\Abf}(\alpp) - \beta\hat{P}^{\alppstarl}_{\Ma'}(\yobs\mid\alppstarl)p_{\Abf}(\alppstarl)\right)\Bigg)^2.
\end{align*}
As detailed in \ref{clt}, with the Central Limit Theorem, we have the following asymptotic confidence level:
\begin{equation}
\gamma(\alpp) = \Phi\left(\frac{\sqrt{\Ma'}\left(\beta\hat{p}^{\alppstarl}_{\Ma'}(\yobs\mid\alppstarl))p_{\Abf}(\alppstarl) -\hat{p}^{\alppstarl}_{\Ma'}(\yobs\mid\alpp)p_{\Abf}(\alpp)\right)}{s_{L'}(\alpp)}\right),
\end{equation}
where $\Phi$ is the cumulative distribution function (cdf) of a Gaussian distribution with mean $0$ and variance $1$, and $s_{L'}(\alpp)$, $\hat{p}^{\alppstarl}_{\Ma'}(\yobs\mid\alppstarl)$ and $\hat{p}^{\alppstarl}_{\Ma'}(\yobs\mid\alpp)$ are the obtained realizations of the associated estimators. 

\subsection{Sampling from the posterior distribution of $\Abf$}\label{full_bayes}

Instead of working only with $\alpmap$ as in the plug-in strategy, we can conduct a full-Bayesian strategy and work with the entire posterior distribution $p(\alpp\mid \yobs)$ estimated by 
\begin{equation}
\hat{P}^{\alppstarl}_{\Ma}(\alpp \mid \yobs) = \frac{\hat{P}^{\alppstarl}_{\Ma}(\yobs\mid\alpp)p_{\Abf}(\alpp)}{\int_{\mathcal{A}} \hat{P}^{\alppstarl}_{\Ma}(\yobs\mid\alpp')p_{\Abf}(\alpp')d\alpp'}. 
\end{equation}

The method is summarized in \Cref{alg-fullbayes}. Note that the first step of this procedure is solely used to estimate the likelihood for all values of $\alpp$, with the estimator $\hat{P}^{\alpp^{\star}_{\ell}}_{\Ma}(\alpp \mid \yobs)$.  Since $\normx{\alpp^{\star}_{\ell} - \alpp^{\star}_{\ell-1}} \leq \thresh$ with small $\thresh$, in practice $\hat{P}^{\alpp^{\star}_{\ell-1}}_{\Ma}(\alpp \mid \yobs)$, obtained from \Cref{alg-map}, can be used to estimate $p(\alpp\mid \yobs)$.  Therefore, no additionnal call to the simulator is required compared to the plug-in method. To simplify the notations and ensure everything depends only on $ \alpp^{\star}_{\ell}$, the method is presented with $\hat{P}^{\alppstarl}_{\Ma}(\alpp \mid \yobs)$ as the estimator.

We have 
\begin{equation}\label{eq_total}
\mathbb{E}(\psi(Y_t(\x_0)) \mid \yobs) = \int_{\mathcal{A}} \mathbb{E}(\psi(Y_t(\x_0))\mid\alpp, \yobs)p(\alpp\mid \yobs)d\alpp.
\end{equation}

As shown in the supplementary material~\cite{charliesire_2024}, under additionnal assumptions on $\lambd \mapsto p(\yobs\mid \lambd)$, $p_\Lambd(.\mid \alpp)$ and $p_\Lambd(\lambd \mid .)$, \\$\int_{\mathcal{A}} \mathbb{E}(\psi(Y_t(\x_0))\mid\alpp, \yobs)\hat{P}^{\alppstarl}_{\Ma}(\alpp \mid \yobs)d\alpp$ is a consistent estimator of $\mathbb{E}(\psi(Y_t(\x_0)) \mid \yobs).$

Therefore, with MCMC methods, we sample $(\Abf_i)_{i=1}^N$ with pdf proportional to
$\hat{P}^{\alppstarl}_{\Ma}(\yobs|.) p_{\Abf}$ and investigate 
\begin{equation}
\tilde{\mathbb{E}}_N^{\alppstarl}(\psi(Y_t(\x_0))) = \frac{1}{N}\sum_{i=1}^N \mathbb{E}(\psi(Y_t(\x_0))\mid\Abf_i,\yobs),
\end{equation}
that approximates $\mathbb{E}(\psi(Y_t(\x_0)) \mid \yobs)$, but is not directly an estimator of $\mathbb{E}(\psi(Y_t(\x_0)) \mid \yobs)$, as $\mathbb{E}(\psi(Y_t(\x_0))\mid\Abf_i,\yobs)$ is unknown. 

Then $\mathbb{E}(\psi(Y_t(\x_0))\mid\Abf_i,\yobs)$ needs to be estimated. As explained in \Cref{hier}, importance sampling estimation can be employed to reduce the computation cost. 
Indeed, from \Cref{eq_post}, we have $\forall \alpp$ s.t. $ \text{supp}(p_{ \Lambd}(.\mid \alpp)) \subset \text{supp}(p_{ \Lambd}(.\mid \alppstarl)),$
\begin{equation}
\mathbb{E}(\psi(Y_t(\x_0))\mid\alpp,\yobs) 
=\frac{\int_{\mathbb{R}^q}\psi(f_t(\x_0, \lambd))p(\yobs \mid \lambd)\frac{p_{ \Lambd}(\lambd\mid \alpp)}{p_{ \Lambd}(\lambd\mid \alppstarl)}p_{ \Lambd}(\lambd\mid \alppstarl)d\lambd}{\int_{\mathbb{R}^q}p(\yobs \mid \lambd)\frac{p_{ \Lambd}(\lambd\mid \alpp)}{p_{ \Lambd}(\lambd\mid \alppstarl)}p_{ \Lambd}(\lambd\mid \alppstarl)d\lambd}.
\end{equation}
This equation gives a full-Bayesian estimator of $\mathbb{E}(\psi(Y_t(\x_0)) \mid \yobs):$
\begin{equation}\label{fullbayes_est}
\ENM{\psi(Y_t(\x_0))}  = \frac{1}{N}\sum_{i=1}^N \frac{\sum_{k=1}^M \psi(f_t(\x_0, \Lambd_k))\frac{p_{ \Lambd}(\Lambd_k\mid \Abf_i)}{p_{ \Lambd}(\Lambd_k\mid \alppstarl)}}{\sum_{k=1}^M \frac{p_{ \Lambd}(\Lambd_k\mid \Abf_i)}{p_{ \Lambd}(\Lambd_k\mid \alppstarl)}},
\end{equation}
with $(\Abf_i)_{i=1}^N$ sampled with pdf proportional to $\hat{P}^{\alppstarl}_{\Ma}(\yobs\mid\alpp)p_{\Abf}(\alpp)$ and $(\Lambd_k)_{k=1}^M$ sampled with pdf proportional to $p(\yobs \mid \lambd)p_{\Lambd}(\lambd\mid \alppstarl).$ To ensure for all $\Abf_i,$ the inclusion $ \text{supp}(p_{ \Lambd}(.\mid \Abf_i)) \subset \text{supp}(p_{ \Lambd}(.\mid \alppstarl))$, $\sum_{k=1}^M \frac{p_{ \Lambd}(\Lambd_k\mid \Abf_i)}{p_{ \Lambd}(\Lambd_k\mid \alppstarl)} > 0,$ and the consistency of the estimators (see Supplementary Material~\cite{charliesire_2024}), in the following we impose 
\begin{equation}\label{supports_eq}
\exists  \text{ compact set } \Kset , \forall \alpp \in \mathcal{A}, \text{supp}(p_{ \Lambd}(.\mid \alpp)) = \Kset.
\end{equation}

\begin{algorithm}[H]
\textbf{Input:} $\alpp^{\star}_\ell$\\
Sample $(\Lambdprime_k)_{k=1}^{\Ma}$ i.i.d. with pdf $p_{\Lambd}(.\mid \alpp^{\star}_\ell).$ \\
Sample $(\Abf_i)_{i=1}^N$ with pdf 
$\propto \hat{P}^{\alppstarl}_{\Ma}(\yobs|.) p_{\Abf}$ by MCMC, with $\hat{P}^{\alppstarl}_{\Ma}(\yobs|.)$ given by \Cref{estimator_likeli}.\\
Sample $(\Lambd_k)_{k=1}^M$ with pdf $\propto p(\yobs \mid \lambd)p_{\Lambd}(\lambd\mid \alppstarl)$ by MCMC. \\
Compute $\ENM{\psi(Y_t(\x_0))}  = \frac{1}{N}\sum_{i=1}^N \frac{\sum_{k=1}^M \psi(f_t(\x_0, \Lambd_k))\frac{p_{ \Lambd}(\Lambd_k\mid \Abf_i)}{p_{ \Lambd}(\Lambd_k\mid \alppstarl)}}{\sum_{k=1}^M \frac{p_{ \Lambd}(\Lambd_k\mid \Abf_i)}{p_{ \Lambd}(\Lambd_k\mid \alppstarl)}}$ \\
\caption{Full-Bayesian estimation of $\mathbb{E}(\psi(Y_t(\x_0)) \mid \yobs).$}\label{alg-fullbayes}
\end{algorithm}

If we consider $N=1$ and $p_{\Abf} = \delta_{\alppstarl}$ in \Cref{fullbayes_est}, and $p_{\Lambd} = p_{\Lambd}(.\mid \alpp^{\star})$ in \Cref{estim_M}, then $\ENM{\psi(Y_t(\x_0))}$ becomes equivalent to $\EM{\psi(Y_t(\x_0))} $. The full-Bayesian estimator is a generalization of the estimator of \Cref{estim_M}.  Consequently, all equations related to $\ENM{\psi(Y_t(\x_0))}$ in the following remain valid for $\EM{\psi(Y_t(\x_0))}.$

More generally, we will denote 
\begin{equation}\label{eq_general_is}
\ENM{h(\Lambd)} = \frac{1}{N}\sum_{i=1}^N \frac{\sum_{k=1}^M h(\Lambd_k)\frac{p_{ \Lambd}(\Lambd_k\mid \Abf_i)}{p_{ \Lambd}(\Lambd_k\mid \alppstarl)}}{\sum_{k=1}^M \frac{p_{ \Lambd}(\Lambd_k\mid \Abf_i)}{p_{ \Lambd}(\Lambd_k\mid \alppstarl)}},
\end{equation}
for every bounded and continuous function $h$ on $\Kset$, with $(\Abf_i)_{i=1}^N$ sampled with pdf proportional to $\hat{P}^{\alppstarl}_{\Ma}(\yobs\mid\alpp)p_{\Abf}(\alpp)$ and $(\Lambd_k)_{k=1}^M$ sampled with pdf proportional to $p(\yobs \mid \lambd)p_{\Lambd}(\lambd\mid \alppstarl).$ It is the full-Bayesian estimator of $\mathbb{E}(h(\Lambd)\mid \yobs)$ and it will be used with different $h$ functions. 
From Sections 2 and 3 of~\cite{charliesire_2024}, we have:
\begin{itemize}
\item $\int_{\mathcal{A}} \mathbb{E}(h(\Lambd)\mid\alpp, \yobs)\hat{P}^{\alppstarl}_{\Ma}(\alpp \mid \yobs)d\alpp$ is a consistent estimator of $\mathbb{E}(h(\Lambd) \mid \yobs).$
\item Given $\hat{P}^{\alppstarl}_{\Ma}(\alpp \mid \yobs) = \hat{p}^{\alppstarl}_{\Ma}(\alpp \mid \yobs),$ $\ENM{h(\Lambd)}$ is a consistent estimator of  $\int_{\mathcal{A}} \mathbb{E}(h(\Lambd)\mid\alpp, \yobs)\hat{p}^{\alppstarl}_{\Ma}(\alpp \mid \yobs)d\alpp.$
\end{itemize}

\subsection{Surrogate modeling}\label{metamodel}

In all of the above, we considered that we can perform runs of the simulator $\f(\x,\lambd)$ to compute the estimators $\EM{\psi(Y_t(\x_0))}$, $\hat{P}_{\Ma}(\yobs\mid \alpp)$ or $\ENM{\psi(Y_t(\x_0))}$. It is important to note that in the approach presented here, the computation of the likelihood $p(\yobs \mid \lambd) = p_{\eps}\left(\yobs - (f_1(\x_j,\lambd))_{j=1}^n \right)$ requires $n$ runs for a single $\lambd.$
It leads to the following numbers of calls to the simulator:
\begin{itemize}
\item $\Ma\times n\times \ell$ runs for the estimation of $\alpmap$, where $\ell$ is the number of iterations before convergence of \Cref{alg-map}.
\item $\Ma\times n + M \times (n+1)$ runs for the estimation of $\ENM{\psi(Y_t(\x_0))}$. The factor $n+1$ comes from the fact that simulations must be run at $(\x_j)_{j=1}^n$ and at the new point $\x_0.$ If $\hat{p}^{\alpp^{\star}_{\ell-1}}_{\Ma}(\alpp \mid \yobs)$ is used for the likelihood approximation, then only $M \times (n+1)$ runs are needed. 
\end{itemize}
In practice, these computations are not feasible when the physical simulations are costly (for instance, several hours long each) and then surrogate models are required. First, surrogate models are required for the approximation of $\left(f_1(\x_j,\lambd)\right)_{j=1}^n$ to compute the likelihood $p(\yobs \mid \lambd) = p_{\eps}\left(\yobs - (f_1(\x_j,\lambd))_{j=1}^n \right)$ for the MCMC sampling in the posterior distribution of $\Lambd$ and the investigation of $\alppstarl$ (see \Cref{alg-map}). Additionally, a surrogate model is needed for $f_t(\x_0,\lambd)$ to compute the prediction estimator at the new point $\x_0$ for the unobserved output $f_t.$ Therefore, for $\x \in \mathbb{X} \cup \{\x_0\}$ and $1\leq t\leq T$, we propose to build a surrogate model for $f_{t,\x} \colon \lambd \longrightarrow f_t(\x, \lambd)$. 

Gaussian processes (GP) are used here as a surrogate model as they make it possible to quantify the uncertainties, as presented in \cite{williams2006gaussian}. Let us assume that $f_{t,\x}$ is a realization of a Gaussian process $Z_{t,\x}(\lambd)$ with prior mean $\mu_{t,\x}(\lambd)$ and prior covariance kernel $k_{t,\x}(\lambd, \lambd')$.
$Z_{t,\x}$ and $Z_{t',\x'}$ are considered independent if $(t,\x)\neq (t',\x')$. The kernel models the correlation structure between $f_{t,\x}(\lambd)$ and $f_{t,\x}(\lambd')$. In our study, the stationary Matérn 5/2 kernel, see \cite{genton}, is used. This choice of independency between the surrogate models implies that the Gaussian process used for the prediction at $\x_0$ is independent of the processes associated with the observation points $\x_j$.

We consider a design of experiments $\mathbb{L}_{\mathrm{train}} = (\lambd_{1},\dots,\lambd_{\ntrain})$ and the observations \\ $Z_{t,\x}^{\mathrm{train}} = (f_{t,\x}(\lambd_{i}))_{i \in I_{\mathrm{train}}}$ with $I_{\mathrm{train}} = \{1,\dots, \ntrain\}$. $f_{t,\x}$ can be approximated using the distribution of $Z_{t,\x}(\lambd) \mid Z(\mathbb{L}_{\mathrm{train}})= Z_{t,\x}^{\mathrm{train}}$,
which is a Gaussian process with the following mean and covariance: 
\begin{equation*}
  \left\{
\begin{aligned}
 \hat{f}_{t}(\x,\lambd) &= \mu_{t,\x}(\lambd) + k_{t,\x}(\lambd, \mathbb{L}_{\mathrm{train}})k_{t,\x}(\mathbb{L}_{\mathrm{train}},\mathbb{L}_{\mathrm{train}})^{-1}(Z_{t,\x}^{\mathrm{train}}-\mu_{t,\x}(\mathbb{L}_{\mathrm{train}})) \\
  c_{t}(\x,\lambd, \lambd') &= k_{t,\x}(\lambd,\lambd') - k_{t,\x}(\lambd, \mathbb{L}_{\mathrm{train}})k_{t,\x}(\mathbb{L}_{\mathrm{train}},\mathbb{L}_{\mathrm{train}})^{-1}k_{t,\x}(\lambd',\mathbb{L}_{\mathrm{train}})^T
\end{aligned}
\right.,
\end{equation*}
where
$k(\mathbb{L}_{\mathrm{train}},\mathbb{L}_{\mathrm{train}}) = [k_{t,\x}(\lambd_i,\lambd_{j})]_{i,j \in I_{\mathrm{train}}}$ is the covariance matrix and \\$k_{t,\x}(\lambd, \mathbb{L}_{\mathrm{train}}) = [k_{t,\x}(\lambd,\lambd_{i})]_{i \in I_{\mathrm{train}}}$.

The variance is denoted $\varv_t(\x, \lambd) = c_{t}(\x,\lambd, \lambd)$. The model presented in \Cref{eq_bay_err} then becomes:
\begin{equation}
\left\{
\begin{aligned}
    Y_t(\x) &= \hat{f}_t(\x, \Lambd) + \sqrt{\varv_t(\x, \Lambd)}\epsf(\x), ~ \: ~ \: ~ \: ~ \: ~ \: \:\:\: 1\leq t\leq T, ~ \x \in \mathcal{X}\\
    \Yobsj &= Y_1(\x_j) + \epsj, ~ \: ~ \: ~ \: ~ \: ~ \: ~ \:\:~ \: ~ \: ~ \: ~ \: ~ \: \:~ \: ~ \:\: \:\:~ \: ~ \: \: ~ \: ~ \: 1\leq j \leq n
\end{aligned}
\right.,
\end{equation}
with $\epsf(\x) \sim \mathcal{N}(0,1)$,  $\epsf(\x) \ind \epsf(\x')$ if $\x\neq \x'$, and  $\eps = (\epsj)_{j=1}^n$ of pdf $p_{\eps}.$

The pdf of the likelihood of $\lambd$ becomes 
$p_{\epstot}\left((\yobsj - \hat{f}_1(\x_j, \lambd))_{j=1}^{n}\right)$ with 
\begin{equation}
\epstot = \left(\epsj + \sqrt{\varv_{1}\left(\x_j,\lambd\right)}\epsf(\x_j)\right)_{j=1}^n.
\end{equation}

Incorporating the variance of the Gaussian process into the likelihood computation ensures that its uncertainty is accounted for during sampling.
Particularly, in our study we consider that $p_{\eps}$ is the pdf of a Gaussian white noise vector, i.e. $\eps \sim \mathcal{N}(0, \sigma_\varepsilon^2I)$. This gives $\epstot \sim \mathcal{N}\left(0, \sigma_\varepsilon^2I +  \text{diag}\Big(\varv_{1}(\x_j, \lambd)\Big)_{j=1}^n\right),$
and the estimators of the prediction mean and variance (from the law of total variance, see \cite{weiss2005course}) are
\begin{equation}\label{mean_var}
\begin{aligned}
\ENM{Y_t(\x_0)} &= \ENM{\hat{f}_t(\x_0, \Lambd)} \\
\ENM{Y_t(\x_0)^2} - \ENM{Y_t(\x_0)}^2 &=  \ENM{\hat{f}_t(\x_0, \Lambd)^2 + \varv_t(\x_0, \Lambd)} - \ENM{\hat{f}_t(\x_0, \Lambd)}^2
\end{aligned}.
\end{equation}

This approach, which uses a separate Gaussian process (GP) for each $\x$, is well-suited to the current context, as it focuses on situations with a very limited number of observations $n$ and a small number of predictions. Note that it is particularly appropriate in other cases where the control variables $\x_j$ are categorical instead of numerical.
However, other implementations are possible, especially with higher values of $n$, by considering a Gaussian process $Z_{t}(\x, \lambd)$ in the joint space $\mathcal{X}\times \Lcal$, with covariance $c_t(\x,\x',\lambd,\lambd').$  In this case, the computation of the likelihood of $\lambd$ would be modified, as the distribution of $\epstot$ is now given by $\epstot \sim \mathcal{N}\left(0, \sigma_\varepsilon^2I +  \big(c_{t}(\x_{j_1},\x_{j_2},\lambd, \lambd)\big)_{j_1, j_2 = 1}^n\right)$

\section{Application}\label{appli}

The previously introduced methods are tested on an application case well-suited for calibration with unobserved outputs, as presented here.

\subsection{Case description}

\begin{figure}[h]
  \centering
  \includegraphics[width=0.45\linewidth]{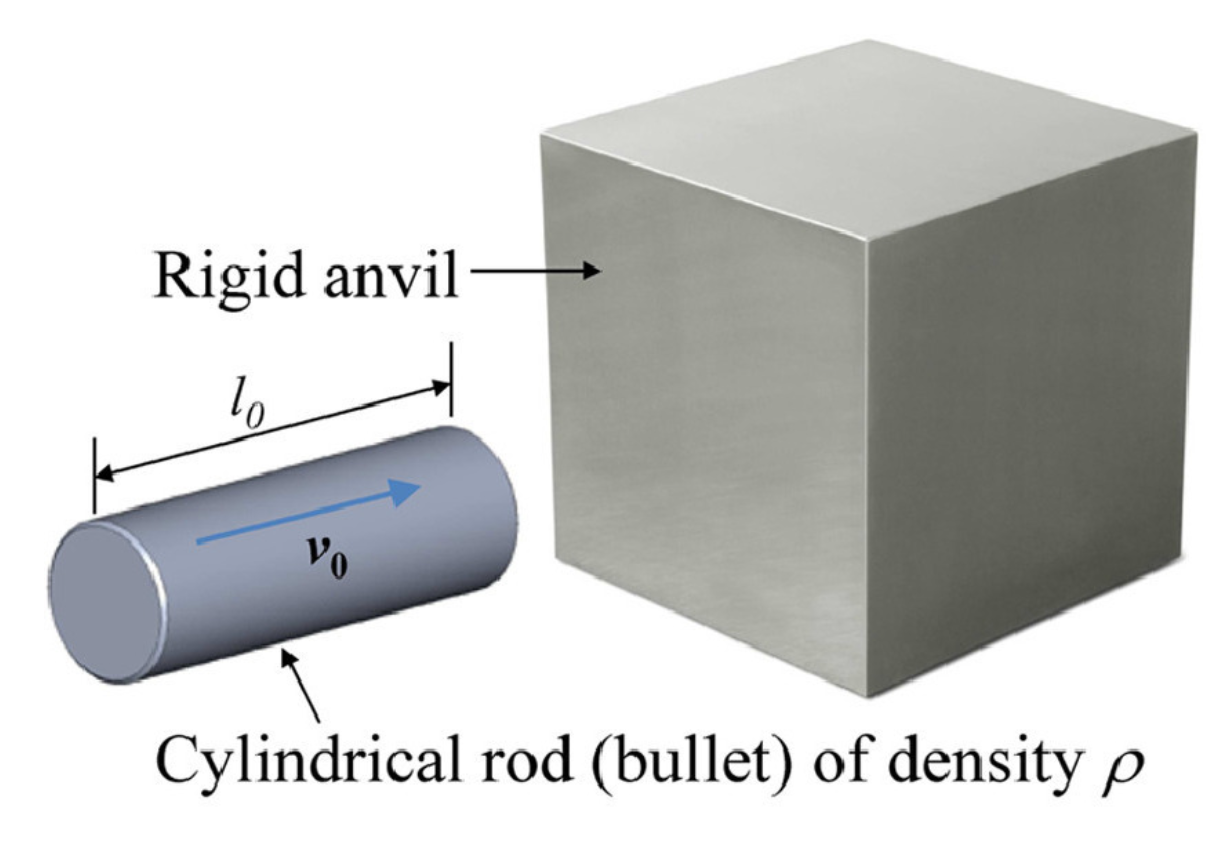}
  \caption{Description of
  %Diagram describing
  the Taylor cylinder impact test on a Copper rod.}
  \label{schema_transpo}
\end{figure}

The test case considered here %was developed by Rémi Chauvin from CEA DAM, and models
deals with the Taylor cylinder impact test \cite{maudlin1999modeling}, as illustrated in \Cref{schema_transpo}. More precisely, it models the impact of a cylinder with length $\ell_0$, radius $r_0$, launched at velocity $v_0$ against a rigid wall. This computer code requires 4 physical parameters associated with the Mie-Grüneisen equation of state \cite{burshtein2008introduction} that are unknown and need to be calibrated. These parameters are $\rho_0$ the intial density of the cylinder, $C_0$ the bulk speed of sound, $\Gamma_0$ the Grüneisen coefficient at the reference state and $S$ the Hugoniot slope parameter. Besides these unknown parameters, the computer code depends on two additional variables, $Y_0$ and $Y_M$, associated with Ludwik’s equation which describes the electroplastic deformation process: $\tau = Y_0 + Y_M \epsilon_p^{n_0}$, where $\tau$ is the stress, $Y_0$ the yield stress, $Y_M$ the strength index, $\epsilon_p$ the plastic strain and $n_0$ is the strain hardening index \cite{mai2011experimental}. 
The two parameters $Y_0$ and $Y_M$ have been calibrated using another experiment, the tensile test, this is why they are considered known here and do not initially require calibration. However, they are reintroduced in the calibration process as additional parameters in $\Lnum$ to capture the model error. As previously explained, the aim is that incorporating this uncertainty will enhance the predictive capability of the model.
Thus, in our study we have :
\begin{itemize}
\item 3 control variables $\x = (\ell_0, r_0,v_0)$
\item $p=4$ parameters to calibrate $\rho_0$, $C_0$, $\Gamma_0$, $S$
\item $q-p = 2$ additional parameters $Y_0$ and $Y_M$. 
\end{itemize}

This computer code, validated on results form \cite{sambasivan2013simulation}, is interesting in our context because it produces $T = 3$ outputs: 
\begin{itemize}
\item The length difference $\ell_f - \ell_{0}$ after impact
\item The final radius $r_f$ 
\item The maximum strain over time $\epsilon_{\text{max}} = \max \:\epsilon_{p}$
\end{itemize}
Therefore, we will consider 3 different transpositions: the first case where we have observations of $\ell_f - \ell_{0}$ only, the second with observations of $r_f$ only, and finally the third with observations of $\epsilon_{\text{max}}$ only. All input variables are constrained by predefined bounds, beyond which the simulation is not feasible. In the following, to simplify, they are all considered as normalized between $0$ and $1$.

\subsection{Model error}

Unfortunately, no real experimental data were available for this study. However, one can easily generate virtual measurements and incorporate a custom model error to work within the previously described framework. This custom model error is introduced through a perturbation of the control variables, ensuring that the synthetic data are generated using a method whose error structure does not exactly replicate that of the inference approach we aim to apply.  
Since our hierarchical methods rely on fluctuations of $Y_0$ and $Y_M$, the model error must not be based on these parameters. Therefore, we choose to perturb the control variables instead.

For a given triplet $\x = (\ell_0, r_0, v_0)$, we define $\perturb(\x) = (\ell_0, r_0, v_0 + \Delta V)$, where $\Delta V$ is fixed to a constant in our work.    
For $1\leq t \leq T$, the true value is introduced as $y_t(\x) = f_t(\perturb(\x), \lambdzero)$ 
for a given $\lambdzero \in \Lcal$, leading to a model error as there is no specific value of $\lambd$ that verifies that $\forall \x \in \mathcal{X},  f_t(\x, \lambd) = y_t(\x).$ 
Then the protocol is the following:
\begin{itemize}
\item Select $n$ points $\mathbb{X} = (\x_j)_{j=1}^n \in  \mathcal{X}^n$ with a Maximin Latin Hypercube Sampling \cite{lhs}. $n$ is small to work with a limited number of observations.
\item Compute $(y_t(\x_j))_{j=1}^n = (f_t(\perturb(\x_j), \lambdzero))_{j=1}^n $,
\item Add a realization $\boldsymbol{\mathcal{E}} = (\mathcal{E}_j)_{j=1}^n$ of the vector $\eps \sim \mathcal{N}(0, \sigma_\varepsilon^2I)$ as the measurement error for the observed variable.
\end{itemize}

Finally it provides the vector of experimental data, $$\yobs = \left(y_1(\x_j) +  \mathcal{E}_j\right)_{j=1}^n.$$

\subsection{Performance metrics}\label{perf_metrics}

Beyond the measurements of the first variable $\yobs = (\yobsj)_{j=1}^n = (y_1(\x_j) + \mathcal{E}_j)_{j=1}^n \in \mathbb{R}^n,$ 
here we also have access to the true values $y_t(\x_j)$ for $1\leq t \leq T, 1\leq j \leq n.$ 
A Leave-One-Out (LOO) approach is used, to predict $y_t(\x_{j})$ for $1 \leq j \leq n$ based on the noisy observations $\yobsloo = (y_1(\x_{j'}) + E_{j'})_{\underset{j'\neq j}{j'=1}}^n$. 
This makes it possible to assess the performance of the method with a small dataset. Then, a specific $\alppstarl$ is identified for every $j$, that will be denoted $\alppstarlj.$ 

More precisely, for $j = 1,\dots,n$ and $t=1,\dots,T,$ we have a random prediction $Y_t(\x_j)\mid  \yobsloo$. Its distribution is called the predictive distribution, that will be compared to the true value $y_t(\x_j).$ In order to have a direct comparison of the different investigated methods, the scoring rule introduced in Equation 25 of \cite{gneiting2007strictly}, based on the mean and the covariance of the predictive distribution, is selected here.
In the general case, it is expressed as 
\begin{equation}\label{score}
s(\boldsymbol{Y}, \boldsymbol{y}) = -\text{log}(\text{det}(\boldsymbol{\Sigma}))- (\boldsymbol{y} - \boldsymbol{\mu})^T\boldsymbol{\Sigma}^{-1}(\boldsymbol{y} - \boldsymbol{\mu}),~~\text{with}
\end{equation}
 
\begin{itemize}
\item $\boldsymbol{Y}$ a random prediction vector of mean $\boldsymbol{\mu}$ and covariance matrix $\boldsymbol{\Sigma}$
\item $\boldsymbol{y}$ the vector of true values of the phenomenon.
\end{itemize}

Here, in the LOO scheme, the predictions $\left(Y_t(\x_j) \mid \yobsloo\right)_{j=1}^{n}$ are independent. The estimated mean and variance of the prediction $Y_t(\x_j) \mid \yobsloo$ are defined as $\hat{\mu}_{t,N,M}(\x_j)$ and $\hat{ \mathrm{v}}_{t,N,M}(\x_j)$, computed using the estimators $\ENMj{Y_t(\x_j)}$ and $ \ENMj{Y_t(\x_j)^2} - \ENM{Y_t(\x_j)}^2$ from \Cref{mean_var}.

For $j = 1,\dots,n$, the score is introduced as $$s_{t,N,M}(\x_j) = -\text{log}(\hat{\mathrm{v}}_{t,N,M}(\x_j)) - \frac{(y_t(\x_j) - \hat{\mu}_{t,N,M}(\x_j))^2}{\hat{\mathrm{v}}_{t,N,M}(\x_j)}
$$
and we finally investigate the overall prediction score
$$s_{t,N,M}(\mathbb{X}) = \sum_{j=1}^{n} s_{t,N,M}(\x_j),$$ which corresponds to the score in \Cref{score} under the assumption of independent predictions, i.e., with $\boldsymbol{\Sigma}$ as a diagonal matrix.
This score provides a meaningful balance between prediction accuracy and uncertainty quantification, making it particularly well-suited to our application. 
The score is computed for all the investigated methods in our study, and the results are compared in \Cref{numerical_results}, with higher scores indicating better predictive performance.

\subsection{Comparison of methods}\label{methods}

To perform a comprehensive analysis of our method, we implement different approaches and compare the previously introduced performance metrics (see \Cref{perf_metrics}). All these approaches are summarized in \Cref{table_recap}. 

\begin{table}[h]
    \centering
    \begin{tabular}{|c|>{\centering\arraybackslash}m{5cm}|>{\centering\arraybackslash}m{5cm}|}
        \cline{2-3}
        \multicolumn{1}{c|}{} & \makecell{\vspace{-1ex} \textcolor{white}{ceci est en blanc}\\\textbf{Formulation}\vspace{2ex} } & \makecell{\textbf{Distribution}} \\
        \hline
      
        \makecell{\\ \textbf{No error}} & \makecell{\vspace{-1ex} \textcolor{white}{ceci est en blanc} \\\vspace{1ex}$Y_t(\x) = f_t(\x, \Lambd)$ \\ 
         $\Lambd \in \mathbb{R}^p$ \vspace{2ex}} & \makecell{\vspace{-1ex} \textcolor{white}{ceci est en blanc} \\\vspace{1ex} $p_\Lambd(\lambd) = \mathbb{1}_{[0,1]^p}$ \vspace{1ex}} \\
        \hline
        \makecell{\\ \textbf{Uniform error}} & \makecell{\vspace{-1ex} \textcolor{white}{ceci est en blanc} \\\vspace{1ex} $Y_t(\x) = f_t(\x, \Lambd)$ \\ $\Lambd \in \mathbb{R}^q$ \vspace{2ex}} & \makecell{\vspace{-1ex} \textcolor{white}{ceci est en blanc} \\\vspace{1ex} $p_\Lambd(\lambd) = \mathbb{1}_{[0,1]^q}$ \vspace{1ex}} \\
        \hline
        \makecell{\\ \\ \textbf{Hierarchical} \\ \textbf{MAP}} & \makecell{\vspace{-1ex} \textcolor{white}{ceci est en blanc} \\\vspace{1ex} $Y_t(\x) = f_t(\x, \Lambd)$ \\ $\Lambd \in \mathbb{R}^q$ \vspace{2ex}} & \makecell{\vspace{-1ex} \textcolor{white}{ceci est en blanc} \\\vspace{1ex} $p_{\Lambd}(\lambd \mid \alppstarl) = $\\$\mathbb{1}_{[0,1]^p}(\lambda_1, \dots, \lambda_p)$\\$\times f_{\mathcal{N}_t(\alppstarl,0.45^2, 0,1)}(\lambda_{p+1}, \dots, \lambda_q)$ \vspace{3ex}} \\
        \hline
        \makecell{\\ \\ \textbf{Hierarchical} \\ \textbf{ full Bayes}} & \makecell{\vspace{-1ex} \textcolor{white}{ceci est en blanc} \\\vspace{1ex} $Y_t(\x) = f_t(\x, \Lambd)$ \\ $\Lambd \in \mathbb{R}^q$ \vspace{2ex}} & \makecell{\vspace{-1ex} \textcolor{white}{ceci est en blanc} \\\vspace{1ex} $p_{\Lambd}(\lambd \mid \Abf) = $\\$\mathbb{1}_{[0,1]^p}(\lambda_1, \dots, \lambda_p)$\\$\times f_{\mathcal{N}_t(\Abf,0.45^2, 0,1)}(\lambda_{p+1}, \dots, \lambda_q)$ \vspace{3ex}} \\
        \hline
        \makecell{\\ \\ \textbf{Embedded} \\ \textbf{discrepancy}} & \makecell{\vspace{-1ex} \textcolor{white}{ceci est en blanc} \\\vspace{1ex} $Y_t(\x) = f_t(\x,\Abf^1 + \delta(\Abf^2, \Xii))$\\ $\Abf^1 \in \mathbb{R}^q, \: \Abf^2 \in \mathbb{R}^q$ \vspace{2ex}} & \makecell{\vspace{-1ex} \textcolor{white}{ceci est en blanc} \\\vspace{1ex}$\delta(\Abf^2, \Xii) = \text{diag}(\Abf^2\Xii^T)$ \\ $\Xii \sim \mathcal{U}([-1,1]^q)$\\ $p_{(\Abf^1, \Abf^2)}(\alpp^1, \alpp^2) \propto$\\$ \prod_{i=1}^{q} \mathbb{1}_{\alpp_i^2 > 0} \mathbb{1}_{\alpp_1^1 - \alpp_i^2 > 0}\mathbb{1}_{\alpp_i^1 + \alpp_i^2 < 1}$ \vspace{3ex}} \\
        \hline
    \end{tabular}
    
    \caption{Summary of all the methods investigated in our study.}
    \label{table_recap}
\end{table}

\FloatBarrier

The first method is to consider no model error in our Bayesian framework, and is called \textit{No error}. We then consider $\lambd = (\rho_0, C_0, \Gamma_0, S) \in \Lphy \subset \mathbb{R}^p$ with a non-informative prior distribution $\mathcal{U}_{[0,1]^p}$. Then we can implement the method described in \Cref{bayes_simple}, and work with the estimator $\EM{\psi(Y_t(\x_0))}$ with $p_{\Lambd} = \mathbb{1}_{[0,1]^p}$.

The second method is to introduce the additionnal parameters $Y_0$ and $Y_M$ to represent the model error, but without a hierarchical model. Indeed, we consider $\lambd = (\rho_0, C_0, \Gamma_0, S, Y_0, Y_M) \in \Lphy\times \Lnum \subset \mathbb{R}^q$ and investigate $\EM{\psi(Y_t(\x_0))}$ with $p_{\Lambd} = \mathbb{1}_{[0,1]^q}.$ It is called \textit{Uniform error}.

The third method, called \textit{Hierarchical MAP}, includes the additionnal parameters $Y_0$ and $Y_M$ with the hierarchical model using a plug-in strategy. It investigates $\alppstarl$ as described in \Cref{alpmap}, and then studies $\EM{\psi(Y_t(\x_0))}$ with $p_\Lambd(\cdot\mid \alppstarl)$. In our work, we consider 

\begin{equation}\label{plambd} 
p_{\Lambd}(\lambd \mid \alpp) = \mathbb{1}_{[0,1]^p}(\lambda_1, \dots, \lambda_p)\times f_{\mathcal{N}_t(\alpp,0.45^2, 0,1)}(\lambda_{p+1}, \dots, \lambda_q),
\end{equation}
where $f_{\mathcal{N}_t(\mu,\sigma^2, a,b)}$ is the pdf of a Gaussian distribution with mean $\mu$ and standard deviation $\sigma$ truncated between $a$ and $b$. This choice of standard deviation of the prior $\sigma = 0.45,$ aims to balance exploratory behavior, allowing the prior distribution to capture model errors, while still concentrating on relevant regions for the additional parameters, in contrast to the uniform error method. Further details on this selection are provided in \ref{study_prior}.

The fourth method is the one described in \Cref{full_bayes}, considering a model error with a hierarchical description but with a full-Bayesian approach, to compute the estimator $\ENM{\psi(Y_t(\x_0))}$. We call it \textit{Hierarchical full Bayes}. We consider the prior $p_{\Lambd} = p_{\Lambd}(\cdot \mid \alpp)$ described in \Cref{plambd}, and $p_{\Abf}$ the pdf of the uniform distribution $\mathcal{U}_{\mathcal{I}(\alppstarl,\alpinter)},$ with $\mathcal{I}(\alppstarl,\alpinter) = [(\alppstarl)_1 - \alpinter, (\alppstarl)_1 + \alpinter]\times \dots \times [(\alppstarl)_r - \alpinter, (\alppstarl)_r + \alpinter]\cap[-10,10]^r,$ and $\alpinter \in \mathbb{R}$. The constant $\alpinter$ is introduced to prevent the investigation of zones with high fluctuations of $\frac{p_{ \Lambd}(\Lambd_k\mid \alpp)}{p_{ \Lambd}(\Lambd_k\mid \alppstarl)}. $  We work with $\alpinter = 4$ here. Note that this choice of prior $p_{\Abf}$ allows for negative values of $\alpp$ (or values higher than $1$), that can concentrate $p_{\Lambd}(\cdot\mid \alpp)$ on a zone near $0$ (or a zone near $1$, respectively).

Finally, the last approach is the one presented in \cite{huan_embedded}, called \textit{Embedded discrepancy} here. Their work is particularly interesting in our context as they consider a model error that is embedded in the calibration parameters and then can be set up in our transposition situation. The method is detailed in \ref{embed}, but it is important to highlight here the difference between the \textit{Embedded discrepancy} and the \textit{Hierarchical full Bayes} approach. 
With our hierarchical model, we consider $Y_t(\x) = f_t(\x, \Lambd),$ where the prior of $\Lambd$ depends on random hyperparameters $\Abf,$ and a Bayesian inference is conducted for both $\Lambd$ and $\Abf$. For a given $\alpp$, the posterior distribution of $\Lambd$ is proportionnal to $p(\yobs\mid \lambd)p_\Lambd(\lambd\mid \alpp).$ However, in the \textit{Embedded discrepancy} approach, the model is $Y_t(\x) = f_t(\x, \Lambd),$ where the distribution of $\Lambd$ depends on random hyperarameters $\Abf = (\Abf^1,\Abf^2).$ Here, the Bayesian inference is conducted only on the hyperparameters $\Abf$. Particularly, for a given $\alpp$, the distribution of $\Lambd$ is directly obtained by $p(\Lambd \mid \alpp)$, without any posterior consideration. The parametrization of $\Lambd \mid \alpp$ plays a crucial role here. Following the recommendations in~\cite{huan_embedded}, we adopt a Legendre-Uniform Polynomial Chaos representation for the distribution of $\Lambd \mid \alpp$, as the parameters $\Lambd$ are bounded here.

Every MCMC sampling is implemented with the package pymcmcstat \cite{Miles2019}, and the optimization of the hyperparameters for the Gaussian Process regression is performed with pylibkriging \cite{libkr}. The same Gaussian Process is used for all the methods. The number of observations is set to $n=10$. $\Ma = 10^4$ samples are considered to estimate $\alpmap$, while $M = 3000$ posterior samples of $\Lambd$ and $N = 750$ samples of $\Abf$ are generated to compute $\ENM{\psi(Y_t(\x_0))}.$ Note that $N< M$ as $\mathcal{A}\subset \mathbb{R}^2$ compared to $\mathcal{L} \subset \mathbb{R}^6$. $\Ma' = 2\times10^4$ samples are used for the estimation of the confidence levels associated with $p(\alpp \mid \yobs) < \beta p(\alppstarl \mid \yobs)$. 
The value of the stopping threshold $\thresh$ in Algorithm~\ref{alg-map} results from a trade-off between choosing a small value to ensure precision and avoiding unnecessary computational cost. In our test case, $\alpp$ corresponds to the mean of a bi-dimensional Gaussian distribution (since $q - p = 2$), truncated between 0 and 1. We set $\tau = 0.02$ considering that below this value, the Euclidean distance between two successive means, $\boldsymbol{\alpha}^\star_\ell$ and $\boldsymbol{\alpha}^\star_{\ell+1}$, is negligible.

\subsection{Numerical results}\label{numerical_results}

The methods are implemented on different designs $\mathbb{X} = (\x_j)_{j=1}^{10}$ for consistency. While only one design is presented here, the remaining results are provided in the "codes/" directory of the Supplementary Material~\cite{charliesire_2024}, and the conclusions remain consistent across all designs. 

\subsubsection{Results overview.}
\begin{figure}[h]
  \centering

  % Première sous-figure
  \begin{subfigure}[b]{1\linewidth}
    \centering
    \includegraphics[width=1\linewidth]{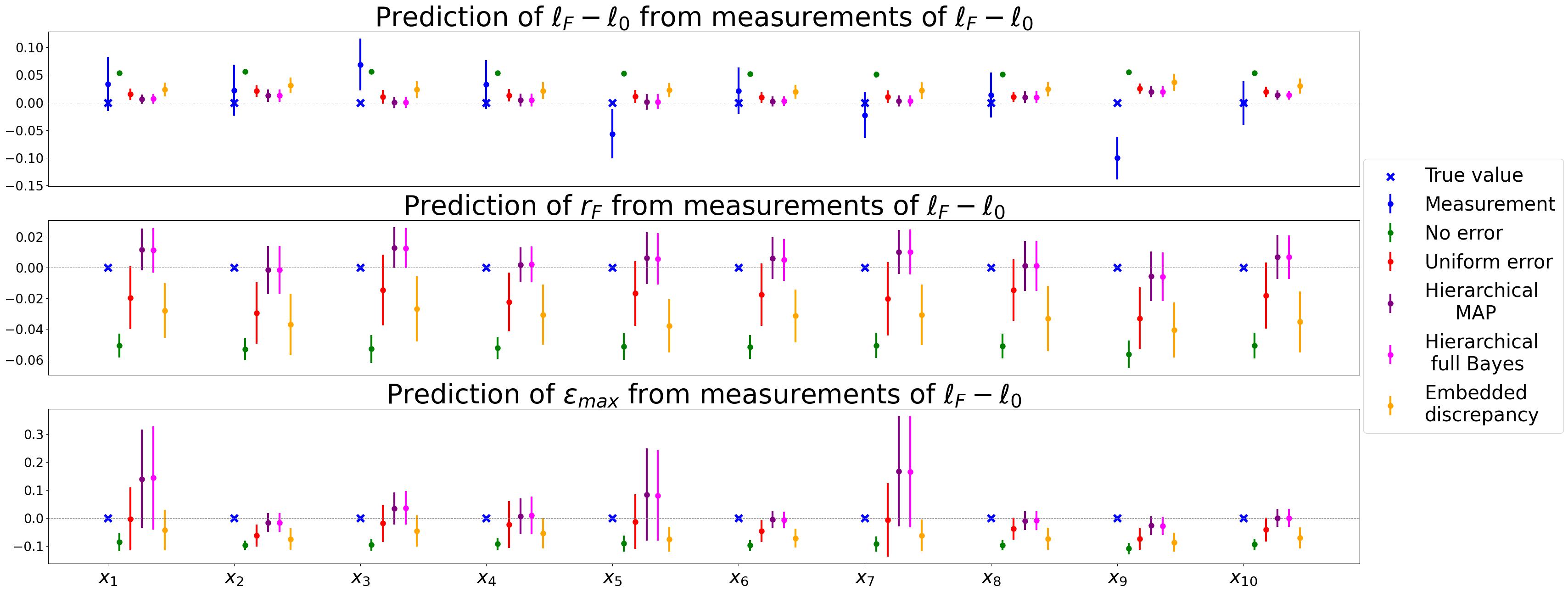}
    \caption{Mean and standard deviation of the normalized posterior distribution $\frac{Y_t(\x_j) - y_t(\x_j)}{y_t(\x_j)}\mid \yobsloo$, $1\leq j \leq 10$, with $y_1 = \ell_f - \ell_{0}$, compared to the true values $y_t(\x_j)$ shifted to $0$ shown as blue crosses for $1 \leq t \leq 3$. The normalized measurement of $\ell_f - \ell_{0}$ is indicated by the blue dot, with its standard deviation represented by the associated error bar.}  % Sous-légende de la première image
    \label{pred_1}
  \end{subfigure}
  
  \vspace{0.5cm} % Espace entre les deux sous-figures
  
  % Deuxième sous-figure
  \begin{subfigure}[b]{1\linewidth}
    \centering
    \renewcommand{\arraystretch}{1.2} % Ajustement de l'espacement vertical
    \begin{tabular}{lccccc}  
        \toprule
        & No error & Uniform  & Hierarchical & Hierarchical & Embedded  \\
       & & error & MAP & full Bayes & discrepancy \\
        \midrule
        Prediction of $\ell_f - \ell_0$ & $-1642.06$ & $6.21$  & $21.40$  & $21.12$  & $-9.88$ \\  
        Prediction of $r_f$ & $-360.11$  & $29.66$ & $45.14$  & $45.29$  & $11.10$  \\  
        Prediction of $\epsilon_{\text{max}}$ & $-137.73$  & $36.52$ & $43.16$  & $42.89$  & $24.41$  \\  
        \bottomrule
    \end{tabular}
    \caption{Scores $s_{t,N,M}(\mathbb{X})$ from observations of $\ell_f - \ell_0$ for each method.}
    \label{tabl_1}
  \end{subfigure}

  \caption{Prediction of the three outputs from observations of $\ell_f - \ell_{0}.$}
  \label{calib_1}

\end{figure}

\begin{figure}[h]
  \centering
  \begin{subfigure}[b]{1\linewidth}
    \centering
    \includegraphics[width=1\linewidth]{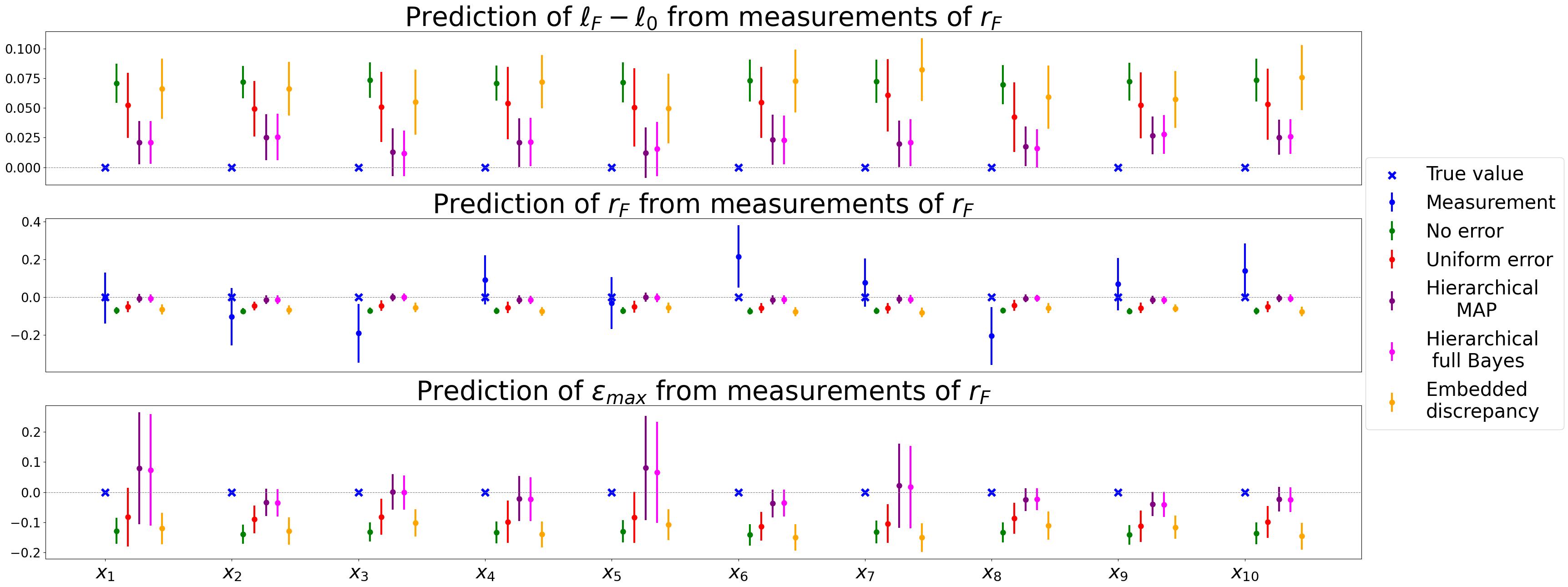}
    \caption{Mean and standard deviation of the normalized posterior distribution $\frac{Y_t(\x_j) - y_t(\x_j)}{y_t(\x_j)}\mid \yobsloo$, $1\leq j \leq 10$, with $y_1 = r_f$, compared to the true values $y_t(\x_j)$ shifted to $0$ shown as blue crosses for $1 \leq t \leq 3$. The measurement of $r_f$ is indicated by the blue dot, with its standard deviation represented by the associated error bar.}  % Sous-légende de la première image
    \label{pred_2}
  \end{subfigure}
  \vspace{0.5cm} 
    \begin{subfigure}[b]{1\linewidth}
      \vspace{0.5cm} 
    \centering
    \renewcommand{\arraystretch}{1.2} %
\begin{tabular}{lccccc}  
        \toprule
        & No error & Uniform  & Hierarchical & Hierarchical & Embedded  \\
        & & error & MAP & full Bayes & discrepancy \\
        \midrule
        Prediction of $\ell_f - \ell_0$ & $-178.16$ & $-22.41$  & $5.13$  & $4.67$  & $-55.48$ \\  
        Prediction of $r_f$ & $-136.48$  & $1.58$ & $37.85$  & $37.89$  & $-39.31$  \\  
        Prediction of $\epsilon_{\text{max}}$ & $-92.16$  & $17.78$ & $40.34$  & $40.68$  & $-29.10$  \\  
        \bottomrule
        
    \end{tabular}
        \caption{Scores $s_{t,N,M}(\mathbb{X})$ from observations of $r_{f}$ for each method.}
    \label{tabl_2}
    \end{subfigure}
  \caption{Prediction of the three outputs from observations of $r_f$}
  \label{calib_2}
  
 \end{figure}

\begin{figure}[h]
  \centering

  % Première sous-figure
  \begin{subfigure}[b]{1\linewidth}
    \centering
    \includegraphics[width=1\linewidth]{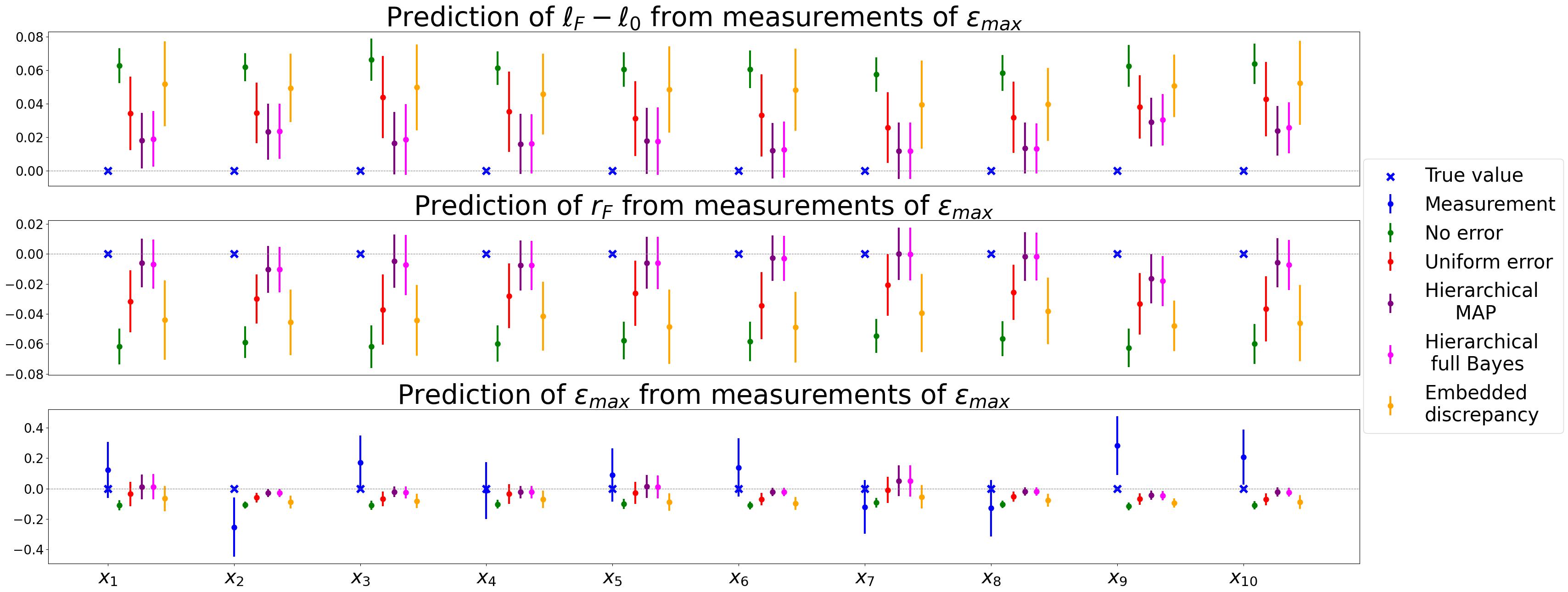}
    \caption{Mean and standard deviation of the normalized posterior distribution $\frac{Y_t(\x_j) - y_t(\x_j)}{y_t(\x_j)}\mid \yobsloo$, $1\leq j \leq 10$, with $y_1 = \epsilon_{\text{max}}$, compared to the true values $y_t(\x_j)$ shifted to $0$ shown as blue crosses for $1 \leq t \leq 3$. The measurement of $\epsilon_{\text{max}}$ is indicated by the blue dot, with its standard deviation represented by the associated error bar.}  % Sous-légende de la première image
    \label{pred_3}
  \end{subfigure}
  
  \vspace{0.5cm} % Espace entre les deux sous-figures
  
  % Deuxième sous-figure
  \begin{subfigure}[b]{1\linewidth}
\centering
    \renewcommand{\arraystretch}{1.2} % Ajustement de l'espacement vertical
    \begin{tabular}{lccccc}  
        \toprule
        & No error & Uniform  & Hierarchical & Hierarchical & Embedded  \\
        & & error & MAP & full Bayes & discrepancy \\
        \midrule
        Prediction of $\ell_f - \ell_0$ & $-305.05$ & $-10.95$  & $7.14$  & $6.39$  & $-28.00$ \\  
        Prediction of $r_f$ & $-182.21$  & $18.66$ & $43.19$  & $42.63$  & $-0.27$  \\  
        Prediction of $\epsilon_{\text{max}}$ & $-92.57$  & $31.77$ & $48.00$  & $47.45$  & $15.53$  \\  
        \bottomrule
    \end{tabular}  % Sous-légende de la deuxième image
      \caption{Scores $s_{t,N,M}(\mathbb{X})$ from observations of $\epsilon_{\text{max}}$ for each method.}
    \label{tabl_3}
  \end{subfigure}

  \caption{Prediction of the three outputs from observations of $\epsilon_{\text{max}}.$}
  \label{calib_3}

\end{figure}

The results are shown in Figures \ref{calib_1}, \ref{calib_2}, \ref{calib_3} corresponding to observations of $\ell_f - \ell_0$, $r_f$ and $\epsilon_{\text{max}}$, respectively. For each figure, the first plot shows the mean and the standard deviation of the normalized posterior distribution $\frac{Y_t(\x_j) - y_t(\x_j)}{y_t(\x_j)}\mid \yobsloo$ for each method and $1\leq j \leq 10$, and compare it to the true values $y_t(\x_j)$ shifted to $0$ represented with blue crosses for $1 \leq t \leq 3$ (see Figures \ref{pred_1}, \ref{pred_2} and \ref{pred_3}). From these posterior distributions, the score $s_{t,N,M}$ is evaluated for each output $t$ and summarized in the associated tables (Figures \ref{tabl_1}, \ref{tabl_2} and \ref{tabl_3}). 

The first key observation is the performance improvement gained by incorporating the additional parameters $Y_0$ and $Y_M$ to represent the model error. Indeed, the true values $y_t$ are clear outliers within the prediction distribution of the \textit{No error} approach (depicted in green on the plots), as evidenced by the significantly negative scores $s_{t,N,M}$, which are much lower than those obtained with methods that account for model error.  

Regarding the model error, the two hierarchical approaches (\textit{Hierarchical MAP} and \textit{Hierarchical full Bayes}) yield nearly identical results and significantly outperform the uniform error approach. These methods effectively shift the prediction mean closer to the true values while maintaining a sufficiently accurate variance estimation. 
This is reflected in the prediction intervals  $
[\hat{\mu}_{t,N,M}(\x_j) - 2\sqrt{\hat{\mathrm{v}}_{t,N,M}(\x_j)}, \hat{\mu}_{t,N,M}(\x_j)  + $ $2\sqrt{\hat{\mathrm{v}}_{t,N,M}(\x_j)}]$
which contain the true values $y_{t}(\x_j)$ for all predictions. This conclusion is further supported by the scores associated with these hierarchical approaches, which are consistently higher than those of the other methods.

The embedded discrepancy approach shows less favorable results compared to the hierarchical approach, with true values $y_{\x_j}$ that are often located in the tails of the prediction distributions, as highlighted by the low score (e.g. clear negative scores for the predictions from observations of $r_{f}$).

\begin{figure}[h]
  \centering
\includegraphics[width=1\linewidth]{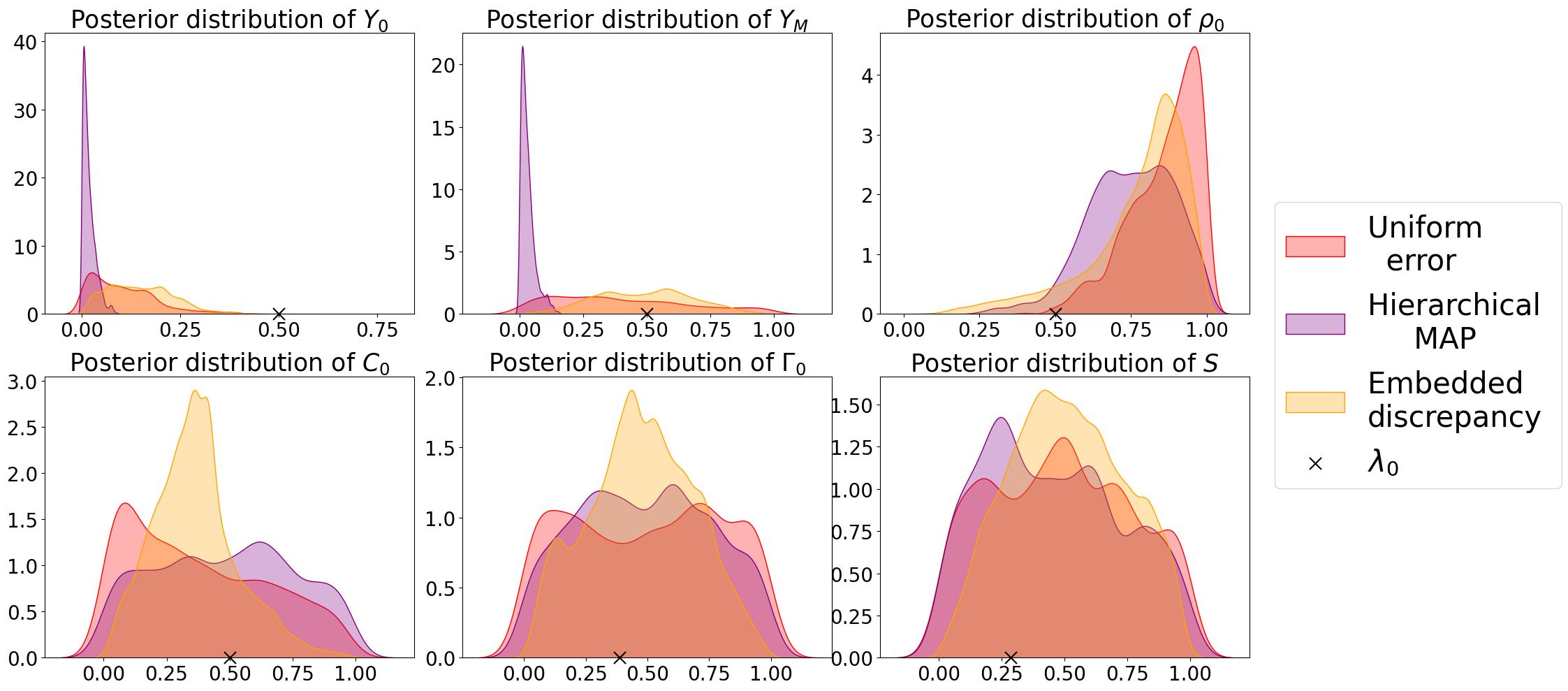}
\caption{Distribution of each marginal of the posterior samples $(\lambd_k)_{k=1}^M$ obtained with the \textit{Uniform error} and the \textit{Hierarchical MAP} strategies, and the samples $\left(\alpp^1_k + \alpp^2_k \xii_r \right)_{\underset{k=1, \ldots, M}{r=1, \ldots, R}}$ for the \textit{Embedded discrepancy}, for the prediction at $\x_{10}$ from observations of $\ell_f - \ell_0$. The black cross is associated with $\lambd_0$, the parameters used for the measurements acquisition. }
  \label{fig_sample}

\end{figure}

To interpret these results, it is necessary to examine $p(\lambd \mid \yobs)$ the pdf of the posterior distribution of $\Lambd$ for the different methods. Since a LOO scheme is applied, a distinct posterior sample $(\lambd_k)_{k=1}^M$ (or $\left(\alpp^1_k + \alpp^2_k \xii_r \right)_{\underset{k=1, \ldots, M}{r=1, \ldots, R}}$ for the \textit{Embedded discrepancy}) is obtained for each $\x_j$ and each method, except for the \textit{Hierarchical MAP} and \textit{Hierarchical full Bayes} methods, where the samples are identical, as discussed in \Cref{full_bayes}. \Cref{fig_sample} presents the marginal distributions of posterior samples obtained for predicting $\x_{10}$ based on measurements of $\ell_f - \ell_0$, but similar results are observed when targeting another $\x_j$ or with a different calibration variable.

\subsubsection{Analysis for the hierarchical model.} 

In \Cref{fig_sample}, it clearly appears that the hierarchical model completely drives the additional parameters $Y_0$ and $Y_M$ to values near 0, which leads to better results. The influence of this approach is amplified by the fact that the observation noise is set high, with standard deviations of $0.9$ for $\ell_f - \ell_0$ and $r_f$, and $0.3$ for $\epsilon_{\text{max}}$ (see Figures \ref{pred_1}, \ref{pred_2}, \ref{pred_3}). Indeed, the larger the noise variance, the less concentrated the likelihood of $\lambd$, thereby increasing the influence of the prior. \ref{appendix_noise} shows the transposition results obtained with smaller observation noises.

The almost identical results between \textit{Hierarchical MAP} and \textit{Hierarchical full Bayes} are explained by the posterior distribution $p(\alpp\mid \yobs)$ that overall focuses on negative values of $\alpp$ and then drives the normalized $\lambd$ very close to $0$, with very small difference compared to the identified $\alppstarl.$ \ref{appendix_alpha} illustrates this posterior distribution for a specific $\x_j$ and $y_1=\epsilon_{\text{max}}$ (see \Cref{likeli_alpha}). It concentrates the prior $p_{\Lambd}(\cdot \mid \alpp)$ on zones near 0 for $\alpp = \alppstarl$ and for all the values $\alpp$ that are sampled with $p(\alpp\mid \yobs)$. \Cref{likeli_alpha} also shows the promising asymptotic confidence levels $\gamma(\alpp)$ associated with $p(\yobs\mid\alpp)p_{\Abf}(\alpp) < 1.05 p(\yobs\mid\alppstarl)p_{\Abf}(\alppstarl),$ as $\forall \alpp, \gamma(\alpp) > 0.99.$

\subsubsection{Analysis for the \textit{embedded discrepancy}.} 
Regarding the performance of the \textit{embedded discrepancy} approach, it is evident that the prediction error consistently occurs in the same direction: $\ell_f - \ell_0$ is overestimated, while $r_f$ and $\epsilon_{\text{max}}$ are underestimated. This can be attributed to the fact that, as is often the case in the presence of model error, the highest likelihoods of $\lambd$ are located at the boundaries, particularly for $Y_0$ and $\rho_0$ (see \Cref{fig_sample}). The influence of these variables on the outputs can be seen as monotonic, as highlighted by linear regression: increasing $Y_0$ leads to an increase in $\ell_f - \ell_0$ but a decrease in $r_f$ and $\epsilon_{\text{max}}$, and the opposite is true for $\rho_0$. However, as expected, the \textit{embedded discrepancy} approach is more exploratory in its sampling, considering a wider range of values for $Y_0$ and $\rho_0$ away from the boundary. This increases the predicted mean of $\ell_f - \ell_0$ and decreases it for $r_f$ and $\epsilon_{\text{max}}$, since no sampling is possible beyond the boundary to balance this exploration.

\section{Summary and perspectives}\label{conclu}

This article addresses the transposition problem, which occurs when observed experimental data do not correspond directly to the data of interest. It examines a scenario with multiple outputs, where only one outputis experimentally observed through a limited set of observations, while the goal is to predict all outputs. The main idea is to propose a representation of the model error that is adapted to this situation, and thus needs to be embedded in the parameters $\Lambd$ to establish a relationship between the observations and the unobserved outputs. Specifically, we explore the inclusion in $\Lambd$ of additional parameters, typically numerical or known physical ones, that were not originally intended for calibration. These parameters are incorporated to account for model error through a hierarchical representation, directing this error into specific zones by introducing hyperparameters $\alpp$ for the prior distribution. The estimation of these hyperparameters is then conducted using Bayesian inference, either by identifying the maximum a posteriori $\alpmap$ or by sampling from the posterior distribution, with an importance sampling scheme to significanlty reduce the computational cost. This study is conducted in the context of expensive simulations, with  Gaussian process regression as a surrogate model. 

The application test case of the article models the Taylor cylinder impact test, which involves three outputs considered as observed variables one at a time, leading to three different sets of results. The performance of our method is compared to the approach proposed by \cite{huan_embedded}, which introduces an \textit{Embedded Discrepancy} method relevant to our scenario. It is important to keep in mind that the results of these Bayesian approaches are highly dependent on the specific application case. 

The results obtained here highlight the significant impact of the hierarchical model, as evidenced by the scores computed to assess the predictive distribution in relation to the true values.
This impact is somewhat reduced when a small noise measurement variance is considered, although the hierarchical approach remains well-suited.  In contrast, the \textit{Embedded Discrepancy} method from \cite{huan_embedded} performs less effectively in our context, partly due to the fact that the most likely values of $\Lambd$ are located near the boundaries. Additionally, the Legendre-Uniform Polynomial Chaos (Legendre-Uniform PC) representation used in this study could be further modified to potentially enhance these results. 

Although the study presented here is comprehensive, further developments should be considered. As explained in \Cref{numerical_results}, the influence of the prior on $\Lambd$ is reduced when the noise measurement variance is small, although the results of the hierarchical model remain promising in this situation, as shown in \ref{appendix_noise}. An in-depth study of the relationship between the impact of the hierarchical representation and the noise measurement could be valuable, potentially identifying a threshold on the noise variance below which the likelihood on $\Lambd$ becomes too concentrated, resulting in a non-influential prior.

Another important aspect of the work is the surrogate model implemented with Gaussian process regression. Here, as explained in \Cref{metamodel}, a different Gaussian process is considered for every $\x \in \mathcal{X}$ and $1\leq t\leq T$, and they are all considered independent, which is possible with a small number of observations. Other implementations are possible. For instance, one could consider a Gaussian process $Z_{t}(\x, \lambd)$ in the joint space $\mathcal{X} \times \Lcal$, or even multi-output Gaussian processes \cite{le2014recursive,alvarez2012kernels,gu2016parallel} to account for the dependencies between the outputs of the computer code $(f_t)_{t=1}^T$.

Even though the framework presented in this article is general, it has only been tested in the context of a transposition situation involving unobserved outputs. Other transposition cases should be explored, such as situations involving a change of scale where predictions are made at a point $\x_0$ far from the observation points $(\x_j)_{j=1}^n.$ Note that the transposition situations can be combined as well, with a change of scale for $\x$ and the prediction of an unobserved output $y_t.$ The method with the hierarchical model and the different performance metrics remain valid in these contexts.

Finally, as previously mentioned, the results of the methods are highly influenced by the application test case. In particular, a computer code without constraints on the calibration variables would be valuable, as it would allow for the consideration of untruncated prior distributions for instance.
\FloatBarrier

\acknowledgements

This research was supported by the consortium in Applied Mathematics CIROQUO, gathering partners in technological and academia in the development of advanced methods for Computer Experiments (see https://doi.org/10.5281/zenodo.6581217). The numerical model used for the Taylor cylinder impact test was developped by Rémi Chauvin.

\section*{Supplementary material}

Supplementary material is available at~\cite{charliesire_2024}. The "proof/" directory contains the proof of convergence for the estimators discussed in Section~\ref{full_bayes}. The "codes/" directory includes the datasets and associated code used to produce the results presented in the article. As noted in Section~\ref{numerical_results}, several experimental designs $\mathbb{X} = (\x_j)_{j=1}^{10}$ were investigated, with results organized into subdirectories named "design\_".
A readme.md file provides additional details about the structure of the repository.

\bibliographystyle{IJ4UQ_Bibliography_Style}

\bibliography{References}

\appendix

\section{Estimation of the confidence levels $\gamma(\alpp)$}\label{clt}

This appendix complements \Cref{conf_in_est} by examining the confidence level \( \gamma(\alpp) \) associated with the condition  
$p(\alpp \mid \yobs) < \beta p(\alppstarl \mid \yobs)$
for a given \( \alpp \), where \( \beta \) represents a user-defined acceptable margin of error for this estimation. The objective is to prove the asymptotic formulation for \( \gamma(\alpp) \) introduced in \Cref{conf_in_est}, using the law of large numbers and a statistical test.
 
$\gamma(\alpp)$ is equivalent to a confidence level for $p(\yobs\mid\alpp)p_{\Abf}(\alpp) <\beta p(\yobs\mid\alppstarl)p_{\Abf}(\alppstarl) $.
We introduce:
$$
\left\{
\begin{aligned}
\mu(\alpp) &= p(\yobs\mid\alpp)p_{\Abf}(\alpp)  - \beta p(\yobs\mid\alppstarl)p_{\Abf}(\alppstarl) \\
v_\beta(\alpp) &=
\mathbb{V}_{\alppstarl}\left( p(\yobs \mid \Lambd)\frac{p_{\Lambd}(\Lambd\mid \alpp)p_{\Abf}(\alpp) - \beta p_{\Lambd}(\Lambd\mid \alppstarl)p_{\Abf}(\alppstarl)}{p_{\Lambd}(\Lambd\mid \alppstarl)}\right), 
\end{aligned}
\right.
$$
where $\Lambd$ has pdf $p_{\Lambd}(. \mid \alppstarl),$ and
\begin{align*}
S_{L'}(\alpp)^2 = &\frac{1}{L'-1}\sum_{k=1}^{L'} \Bigg( p(\yobs \mid \Lambd'_k)\frac{p_{\Lambd}(\Lambd'_k\mid \alpp)p_{\Abf}(\alpp) - \beta p_{\Lambd}(\Lambd'_k\mid \alppstarl)p_{\Abf}(\alppstarl)}{p_{\Lambd}(\Lambd'_k\mid \alppstarl)}- \\
&\left(\hat{P}^{\alppstarl}_{\Ma'}(\yobs\mid\alpp)p_{\Abf}(\alpp) - \beta\hat{P}^{\alppstarl}_{\Ma'}(\yobs\mid\alppstarl)p_{\Abf}(\alppstarl)\right)\Bigg)^2,
\end{align*}
where $(\Lambdprime_k)_{k=1}^{\Ma'}$ are i.i.d. with pdf $p_{\Lambd}(. \mid \alppstarl).$
From the law of large numbers, $S_{L'}(\alpp)^2 \xrightarrow[L'\to\infty]{p} v_\beta(\alpp)$ and then $\frac{S_{L'}(\alpp)}{\sqrt{v_\beta(\alpp)}} \xrightarrow[L'\to\infty]{p} 1.$
The Central Limit Theorem \cite{Korolev} provides that, given $\alppstarl$, with $\hat{P}^{\alppstarl}_{\Ma'}(\yobs\mid\alpp)$ defined by \Cref{estimator_likeli}, 
$$
\frac{\sqrt{L'}\Bigg(\hat{P}^{\alppstarl}_{\Ma'}(\yobs\mid\alpp)p_{\Abf}(\alpp) - \beta\hat{P}^{\alppstarl}_{\Ma'}(\yobs\mid\alppstarl)p_{\Abf}(\alppstarl) - \mu(\alpp)\Bigg)}{ \sqrt{v_\beta(\alpp)}} \xrightarrow[L'\to\infty]{d}\mathcal{N}(0,1).
$$
Slutsky's theorem then provides 
\begin{equation}\label{slusky}
\frac{\sqrt{L'}\Bigg(\hat{P}^{\alppstarl}_{\Ma'}(\yobs\mid\alpp)p_{\Abf}(\alpp) - \beta\hat{P}^{\alppstarl}_{\Ma'}(\yobs\mid\alppstarl)p_{\Abf}(\alppstarl) - \mu(\alpp)\Bigg)}{S_{L'}(\alpp)} \xrightarrow[L'\to\infty]{d}\mathcal{N}(0,1).
\end{equation}

We introduce the hypothesis of a statistical test $H_0\: :\: \mu(\alpp) \geq 0.$ We consider $\gamma(\alpp)$ the complementary of the associated p-value, defined by 
\begin{align*}
\gamma(\alpp) = 1 - \mathbb{P}\Biggl[&\frac{\hat{P}^{\alppstarl}_{\Ma'}(\yobs\mid\alpp)p_{\Abf}(\alpp) - \beta\hat{P}^{\alppstarl}_{\Ma'}(\yobs\mid\alppstarl)p_{\Abf}(\alppstarl)}{S_{L'}(\alpp)} \leq \\
&\frac{\hat{p}^{\alppstarl}_{\Ma'}(\yobs\mid\alpp)p_{\Abf}(\alpp) - \beta\hat{p}^{\alppstarl}_{\Ma'}(\yobs\mid\alppstarl)p_{\Abf}(\alppstarl)}{s_{L'}(\alpp)}\Biggr]
\end{align*} under $\mu(\alpp) = 0$, with $s_{L'}(\alpp)$, $\hat{p}^{\alppstarl}_{\Ma'}(\yobs\mid\alppstarl)$ and  $\hat{p}^{\alppstarl}_{\Ma'}(\yobs\mid\alpp)$ the obtained realizations of the associated estimators. Then, $\gamma(\alpp)$ is indeed a confidence level associated with $\mu(\alpp)<0$, the alternative hypothesis. 
Under $\mu(\alpp) = 0$, with Equation \ref{slusky}, we have for $z \in \mathbb{R},$
$$\mathbb{P}\biggl[\frac{\hat{P}^{\alppstarl}_{\Ma'}(\yobs\mid\alpp)p_{\Abf}(\alpp) - \beta\hat{P}^{\alppstarl}_{\Ma'}(\yobs\mid\alppstarl)p_{\Abf}(\alppstarl)}{S_{L'}(\alpp)} \leq \frac{z}{\sqrt{L'}}\biggr]  \xrightarrow[L'\to\infty]{} \Phi(z).$$

with $\Phi$ is the cdf of a Gaussian distribution with mean $0$ and variance $1$.

Finally, with $z = \frac{\sqrt{L'}\left(\hat{p}^{\alppstarl}_{\Ma'}(\yobs\mid\alpp)p_{\Abf}(\alpp)-\beta\hat{p}^{\alppstarl}_{\Ma'}(\yobs\mid\alppstarl)p_{\Abf}(\alppstarl) \right)}{s_{L'}(\alpp)},$ we have the following asymptotic approximation for the confidence level

\begin{equation}
\gamma(\alpp) = \Phi\left(\frac{\sqrt{\Ma'}\left(\beta\hat{p}^{\alppstarl}_{\Ma'}(\yobs\mid\alppstarl))p_{\Abf}(\alppstarl) -\hat{p}^{\alppstarl}_{\Ma'}(\yobs\mid\alpp)p_{\Abf}(\alpp)\right)}{s_{L'}(\alpp)}\right).
\end{equation}

\section{Prior fluctuations of the outputs}\label{study_prior}

As mentioned in \Cref{intro}, the fluctuations of the additional parameters must introduce sufficient variations in the outputs to adequately capture the model error. The objective here is to investigate, for our application test case, the amplitude of the variations in the three outputs when the calibration parameters \( \Lambd \) vary according to their prior distribution. This analysis aims to confirm the applicability of our strategy in this context and to provide insights into the appropriate standard deviation for the prior distribution.

\begin{figure}[h]
    \centering
    \includegraphics[width=\linewidth]{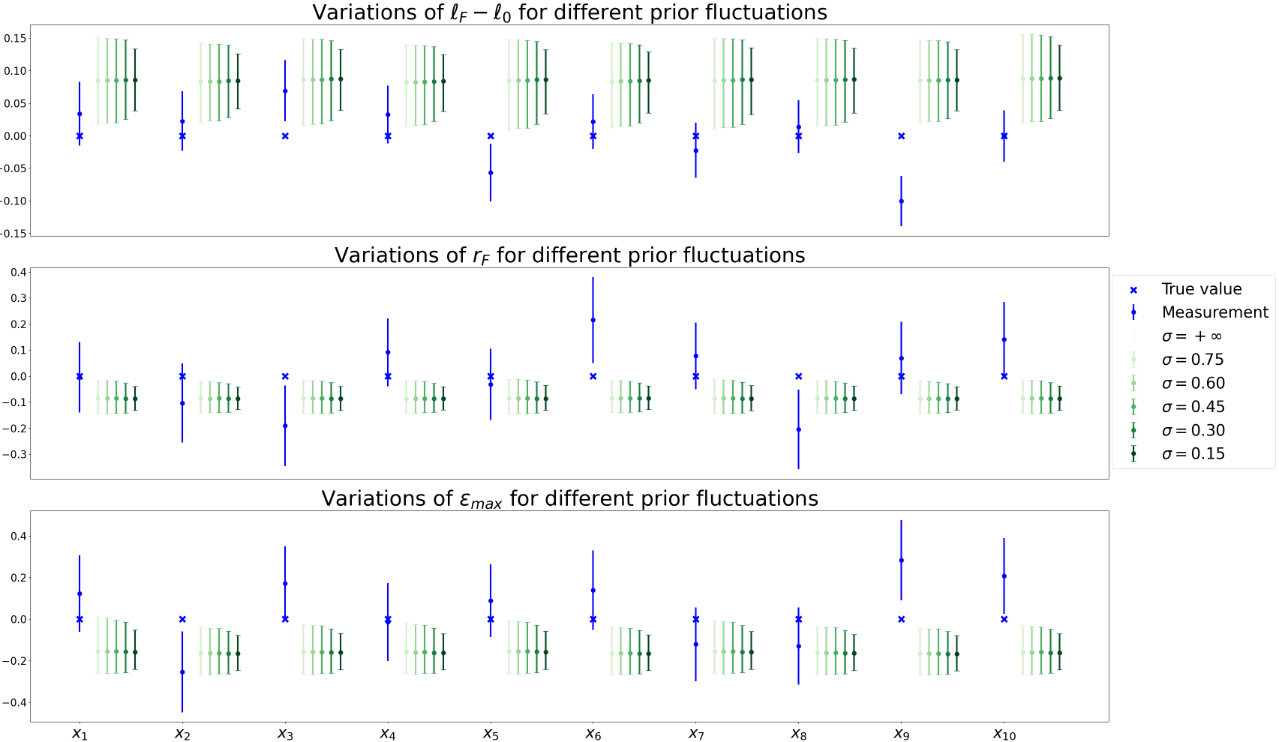}
    \caption{Variations in the output variables due to the prior fluctuations of the parameters $\Lambd$.}
    \label{prior_variations}
\end{figure}

In Figure~\ref{prior_variations}, these variations are analyzed for prior distributions with different standard deviations. The idea is to visualize the distribution of \( f_t(\x, \Lambd) \) where $\Lambd$ has density $p^{\sigma}_{\Lambd}(\lambd) = \mathbb{1}_{[0,1]^p}(\lambda_1, \dots, \lambda_p)\times f_{\mathcal{N}_t(0.5,\sigma^2, 0,1)}(\lambda_{p+1}, \dots, \lambda_q)$ for different values of \( \sigma \), where \( \mathcal{N}_t(0.5, \sigma^2, 0,1) \) is a Gaussian distribution with mean \( 0.5 \) and standard deviation \( \sigma \), truncated between \( 0 \) and \( 1 \).  More precisely,  i.i.d. samples \( \left(\Lambd_{k}\right)_{k=1}^{10^4} \) are generated with density $p^{\sigma}_{\Lambd}(\lambd)$, and the interval \([y_{t,\sigma}^{0.025}(\x), y_{t,\sigma}^{0.975}(\x)]\) is displayed to study the fluctuations of the outputs, where \( y_{t,\sigma}^{0.025}(\x) \) and \( y_{t,\sigma}^{0.975}(\x) \) are the empirical quantiles of level \( 0.025 \) and \( 0.975 \) of the sample \( \left(f_{t}(\x, \Lambd_k)\right)_{k=1}^{10^4} \).  

The objective is to identify standard deviations that are too small to capture the model discrepancy when using centered prior distributions. In Figure~\ref{prior_variations}, the case $\sigma = +\infty$ corresponds to uniform sampling in $[0,1]^q$, which is naturally the most exploratory scenario. This confirms that our strategy is applicable here, as the variations in the outputs are large enough to account for the model discrepancy.  

Moreover, the exploration achieved with uniform sampling appears nearly identical to that obtained with prior distributions using $\sigma = 0.75$, $\sigma = 0.60$, and $\sigma = 0.45$. In contrast, smaller values such as $\sigma = 0.30$ and $\sigma = 0.15$ lead to reduced output variations, which do not allow for generating simulations sufficiently close to the true values of the phenomenon.  

To further investigate the choice of $\sigma$, we introduce the function  
\[
g^j_t(\sigma) = \underset{y \in [y_{t,\sigma}^{0.025}(\x_j), y_{t,\sigma}^{0.975}(\x_j)]}{\text{inf}} \lvert y_t(\x_j) - y \rvert, \quad \text{for } 1 \leq j \leq n.
\]  
This function measures how the variations in the output $f_t(\x_j, \Lambd)$ allow for getting closer to the true value $y_t(\x_j)$ when $\Lambd$ is sampled according to the density $p^{\sigma}_{\Lambd}$. To obtain a global metric, we define the aggregated function:  
\[
g_t(\sigma) = \frac{1}{n} \sum_{j=1}^{n} g_t^j(\sigma).
\]

\begin{figure}[h]
    \centering
    \includegraphics[width=\linewidth]{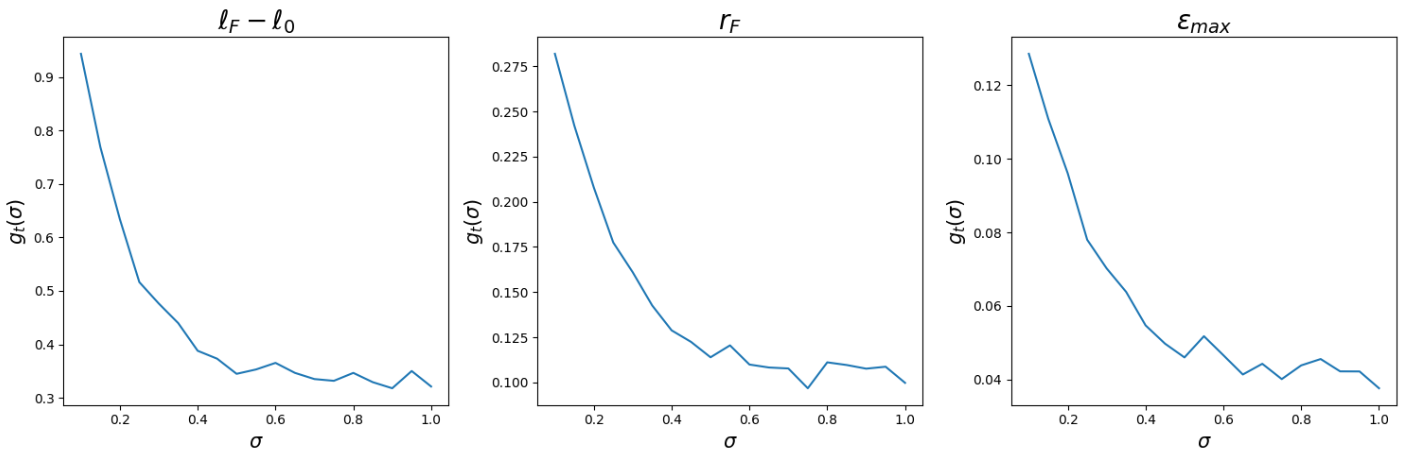}
    \caption{Evolution of $g_t(\sigma)$ for the three outputs.}
    \label{prior_error}
\end{figure}

Analyzing the evolution of $g_t$ as $\sigma$ increases in \Cref{prior_error}, we observe that small values of $\sigma$ do not provide sufficient exploration to approximate the true values of the phenomenon. This results in an initial regime where $g_t$ decreases sharply until $\sigma \approx 0.45$. Beyond this threshold, the gain in exploration becomes less significant, leading to a regime close to a plateau. This observation supports the choice of a standard deviation around $\sigma = 0.45$. Note that the function $g_t$ is not monotonic here due to sampling errors.

\section{Embedded discrepancy}\label{embed}

This section provides a brief summary of the work by \cite{huan_embedded}, who proposed another type of embedded error applicable in our transposition context. Since this method has been implemented in our work for comparison with our numerical results, these details are essential for understanding the fundamentals of their approach.  

Their idea is to consider an embedded random discrepancy $\delta$, parameterized by $\alpp^2$. In the following, this discrepancy is denoted $\delta(\alpp^2, \Xii)$, where $\Xii$ highlights explicitly the stochastic dimension. 

This formulation leads to considering the random output $f_t(\x, \alpp^1 + \delta(\alpp^2, \Xii))$, where $\Lambd = \alpp^1 + \delta(\alpp^2, \Xii)$ is a random variable parametrized by $\alpp^1$ and $\alpp^2$. The problem is treated as a Bayesian estimation of the parameters $\alpp^1$ and $\alpp^2$, that are parameters of a pdf. In the following, we denote $\alpp = (\alpp^1,\alpp^2)$ the set of augmented parameters. Then, we consider the GP formulation
$$Y_t(\x) = \hat{f}_t(\x,\Abf^1 + \delta(\Abf^2, \Xii)) + \sqrt{\varv_{1}\left(\x, \Abf^1 + \delta(\Abf^2, \Xii)\right)}\epsf,$$

with  $\Abf^1 \in \mathbb{R}^q, \: \Abf^2 \in \mathbb{R}^q.$

\subsection{Likelihood estimation.}

We need to investigate $p(\yobs\mid \alpp)$ the likelihood of $\alpp.$ We first define the random vectors

\begin{equation*}
\begin{cases}
\FF(\alpp) = \left( \hat{f}_1(\x_j, \alpp^1 + \delta(\alpp^2, \Xii))\right)_{j=1}^n \\
\HH(\alpp) =\left(  \hat{f}_1(\x_j, \alpp^1 + \delta(\alpp^2, \Xii)) + \sqrt{\varv_{1}(\x, \alpp^1 + \delta(\alpp^2, \Xii))}\epsf + \epsj\right)_{j=1}^n
\end{cases}
\end{equation*}

Then, we have $$p(\yobs \mid \alpp) = \pi_{\HH(\alpp)}(\yobs)$$ with $\pi_{\HH(\alpp)}$ the pdf of $\HH(\alpp)$. One challenge is to evaluate $\pi_{\HH(\alpp)}$ for every $\alpp$. A high dimensional Kernel Density Estimation (KDE, \cite{terrell1992variable}) would provide a precise computation but is too costly in our context. Several likelihood approximations are proposed in \cite{huan_embedded}, and we have decided to work with an independent normal approximation here, for simplicity reasons. The idea is to get $R$ samples $(\xii_r)_{r=1}^{R}$ that are i.i.d. realizations of the random variable $\Xii$, and then compute for $1\leq j \leq n$

\begin{equation*}
\begin{cases}
\hat{\mu}_j &= \frac{1}{R} \sum_{r=1}^{R}  \hat{f}_1(\x_j, \alpp^1 + \delta(\alpp^2,  \xii_r)) \\
\hat{\sigma}_j^2 &= \frac{1}{R-1} \sum_{r=1}^{R} (\hat{f}_1(\x_j, \alpp^1 + \delta(\alpp^2,  \xii_r)) - \hat{\mu}_j)^2 + \frac{1}{R} \sum_{r=1}^R \varv_{1}(\x_j, \alpp^1 + \delta(\alpp^2, \xii_r)) + \sigma_{\varepsilon}^2
\end{cases}
\end{equation*}

Finally, with the independent normal approximation, the pdf $\pi_{\HH(\alpp)}$ is defined as $$\pi_{\HH(\alpp)}(\mathbf{y}) = \prod_{j=1}^{n}\frac{1}{\sqrt{2\pi}\hat{\sigma}_j}exp(-\frac{(y_j - \hat{\mu}_j)^2}{2\hat{\sigma}_j^2})~.$$

\subsection{Polynomial chaos representation.}

The random discrepancy $\delta(\alpp^2, \Xii)$ needs to be parametrized. As presented in \cite{huan_embedded}, we have opted for a Polynomial Chaos (PC, \cite{pce}) representation, and more precisely Legendre-Uniform PC that can respect the support of the calibration parameters. It gives 

$$\alpp^1 + \delta(\alpp^2,\Xii) = \alpp^1 + \text{diag}(\alpp^2\Xii^T)
$$

with $\Xii_i \sim \mathcal{U}(-1,1), ~1\leq i \leq q$ and $\Xii_i \ind \Xii_j$ if $i\neq j.$

\subsection{Prior construction. }

The choice of the prior is here driven by the respect of the support of the parameters once again. We have opted for the uniform prior that respects the constraints $\alpp^1_i - \alpp^2_i \geq 0$ and $\alpp^1_i + \alpp^2_i \leq 1,$ and $\alpp^2_i \geq 0$ as well to avoid bimodal distributions.
We then have  
$$p_{(\Abf^1, \Abf^2)}(\alpp^1, \alpp^2) \propto \prod_{i=1}^{q} \mathbb{1}_{\alpp_i^2 > 0} \mathbb{1}_{\alpp_i^1 - \alpp_i^2 > 0}\mathbb{1}_{\alpp_i^1 + \alpp_i^2 < 1}.$$

\subsection{Computation cost. }

This strategy is relevant in our context but involves a high computation cost compared to the method presented in this article. Indeed, computing the likelihood for a single $\alpp$ requires $R \times n$ calls to the simulator (or the surrogate model), compared to $n$ calls in our method. This is a major difference as $R$ is associated with sampling in dimension $q$ and thus must be large.

\section{Posterior distribution and MAP for $\alpp$}\label{appendix_alpha}

\begin{figure}[h]
  \centering

  \begin{subfigure}[b]{0.45\linewidth}
    \centering
    \includegraphics[width=\linewidth]{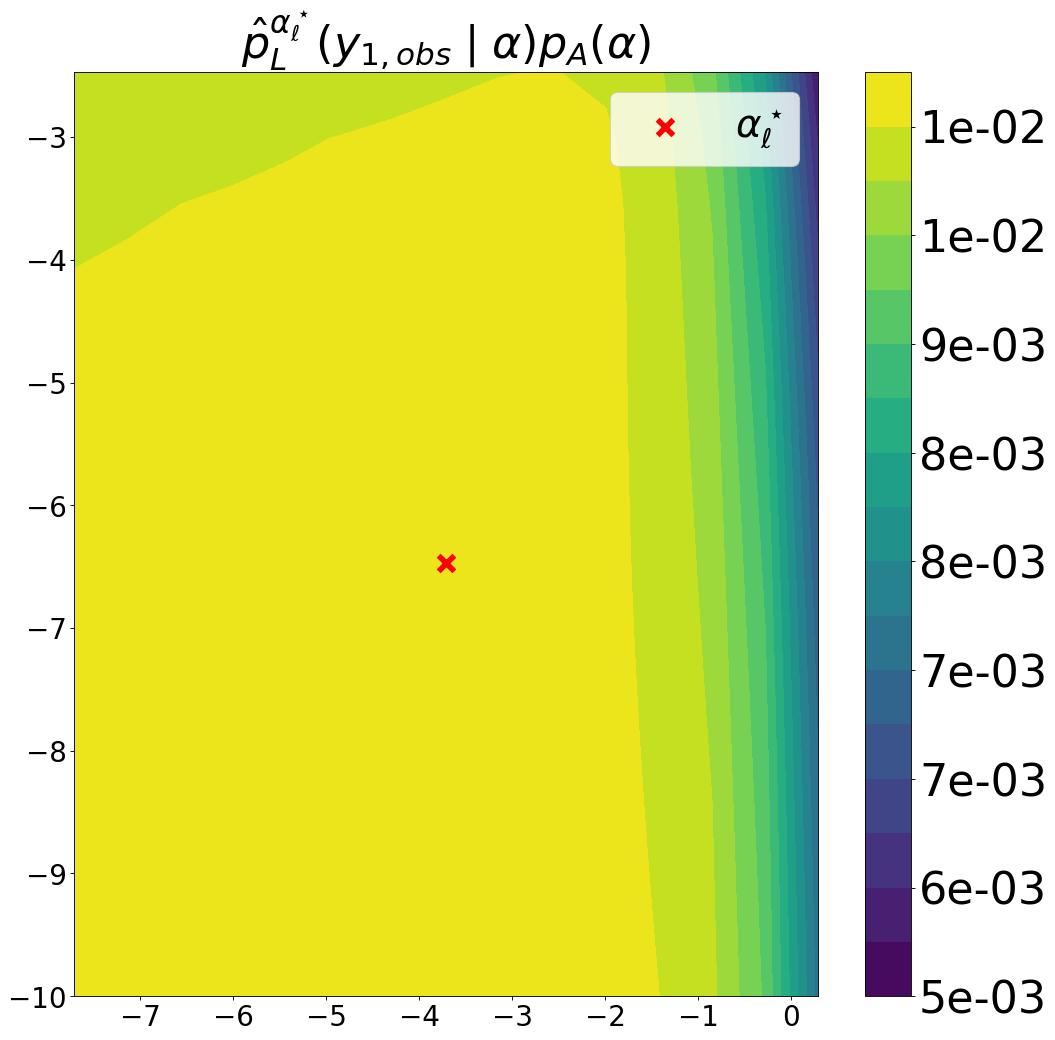}
    \caption{Estimated unnormalized posterior distribution $\hat{p}^{\alppstarlj}_{\Ma}(\yobsloo\mid\alpp)p_{\Abf}(\alpp)$.}
    \label{likeli_alpha}
  \end{subfigure}
  \hfill
  \begin{subfigure}[b]{0.49\linewidth}
    \centering
    \includegraphics[width=0.97\linewidth]{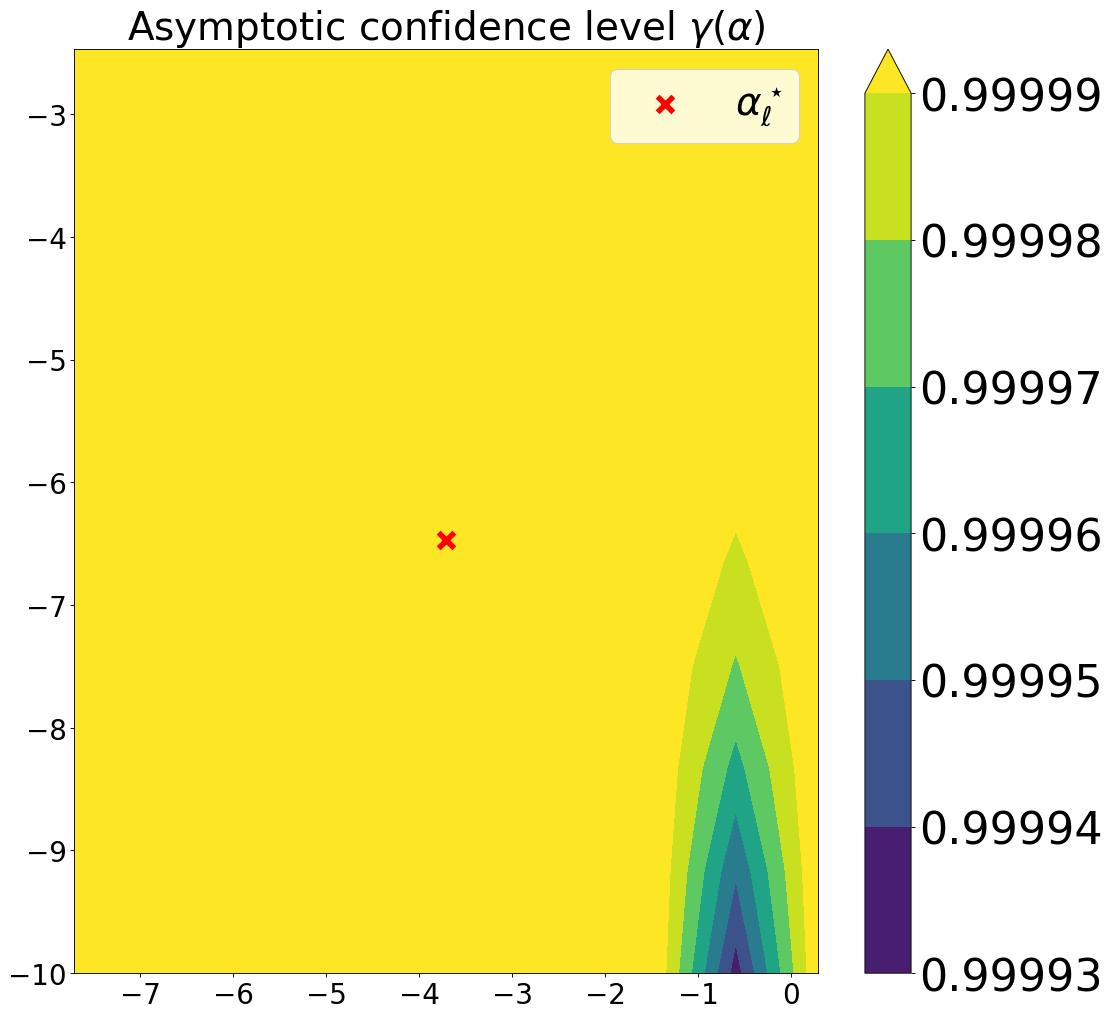}
    \caption{Asymptotic confidence level $\gamma(\alpp)$ associated with $p(\alpp \mid \yobsloo) < 1.05 p(\alppstarlj \mid \yobsloo)$}
    \label{confidence_alpha}
  \end{subfigure}

  \caption{Illustration of the estimation of the posterior distribution $p(\alpp\mid \yobsloo)$ (\Cref{likeli_alpha}) and the confidence level $\gamma(\alpp)$ associated to $p(\alpp \mid \yobsloo) < 1.05 p(\alppstarlj \mid \yobsloo)$ (\Cref{confidence_alpha}) with $y_1 =\epsilon_{\text{max}}$ for the predication at $\x_6$.}
  \label{fig_alpha}
\end{figure}

This appendix provides details about the posterior distribution of $p(\alpp\mid\yobsloo)$ in the application test case. \Cref{likeli_alpha} presents the estimated unnormalized posterior distribution \(\hat{p}^{\alppstarlj}_{\Ma}(\yobsloo\mid\alpp)p_{\Abf}(\alpp)\) for \( j = 6 \), derived from observations of \( \epsilon_{\text{max}} \). It highlights that this distribution is concentrated on negative values of \( \alpp \).  
 \Cref{confidence_alpha} provides the asymptotic confidence levels $\gamma(\alpp)$ associated to $p(\alpp \mid \yobsloo) < 1.05 p(\alppstarlj \mid \yobsloo)$, estimated to validate that the posterior pdf at $\alppstarlj$ is close to the maximum posterior density across all $\alpp \in \mathcal{A}.$ Here, the confidence levels are promising with $\forall \alpp, \gamma(\alpp) > 0.99.$

\FloatBarrier
\section{Example of results with smaller noise}\label{appendix_noise}

This appendix examines a scenario where the observation noise variance is reduced compared to the results shown in \Cref{numerical_results}.

\begin{figure}[h]
  \centering

  % Première sous-figure
  \begin{subfigure}[b]{1\linewidth}
    \centering
    \includegraphics[width=1\linewidth]{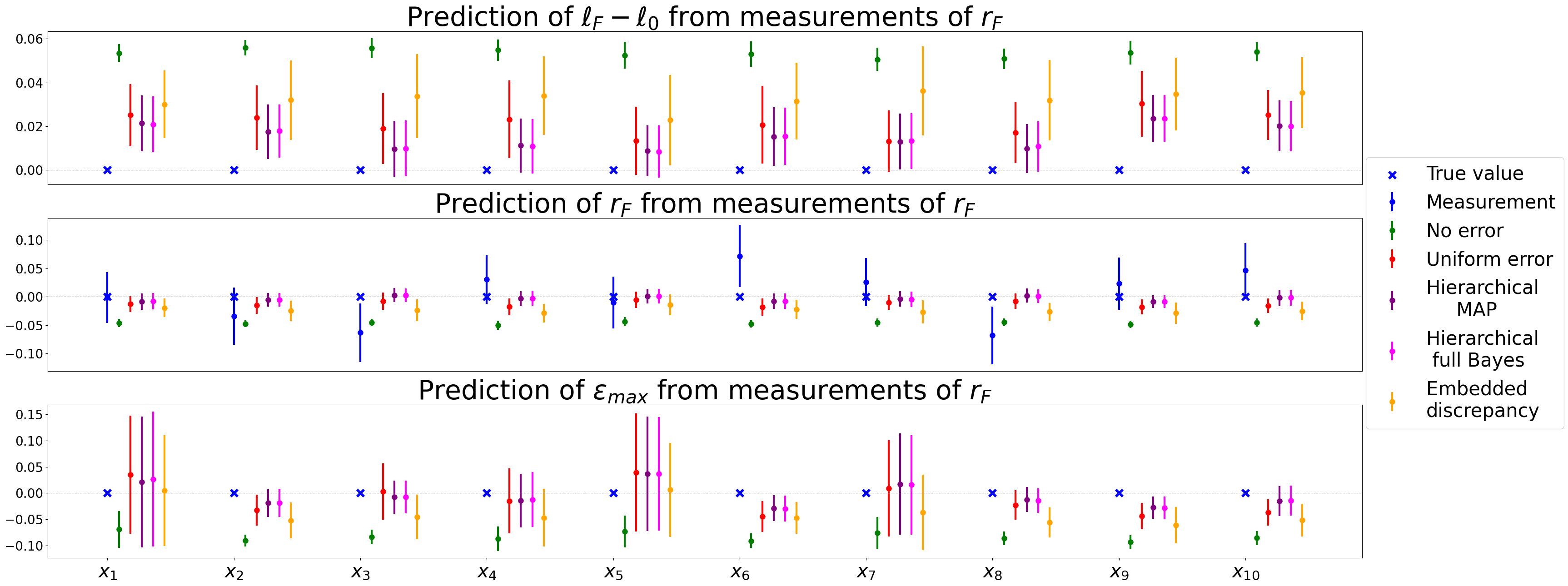}
    \caption{Mean and standard deviation of the normalized posterior distribution $\frac{Y_t(\x_j) - y_t(\x_j)}{y_t(\x_j)}\mid \yobsloo$, $1\leq j \leq 10$, with $y_1 = r_f$, compared to the true values $y_t(\x_j)$ shifted to $0$ shown as blue crosses for $1 \leq t \leq 3$. The measurement of $r_f$ is indicated by the blue dot, with its standard deviation represented by the associated error bar.}  % Sous-légende de la première image
    \label{pred_1noise}
  \end{subfigure}
  
  \vspace{0.5cm} % Espace entre les deux sous-figures
  \begin{subfigure}[b]{1\linewidth}
\centering
    \renewcommand{\arraystretch}{1.2} % Ajustement de l'espacement vertical
    \begin{tabular}{lccccc}  
        \toprule
        & No error & Uniform  & Hierarchical & Hierarchical & Embedded  \\
        & & error & MAP & full Bayes & discrepancy \\
        \midrule
        Prediction of $\ell_f - \ell_0$ & $-1298.05$ & $1.15$  & $9.72$  & $9.47$  & $-13.62$ \\  
        Prediction of $r_f$ & $-385.63$  & $38.98$ & $48.55$  & $48.61$  & $24.42$  \\  
        Prediction of $\epsilon_{\text{max}}$ & $-235.49$  & $41.72$ & $49.02$  & $48.85$  & $34.97$  \\  
        \bottomrule
    \end{tabular}  % Sous-légende de la deuxième image
      \caption{Scores $s_{t,N,M}(\mathbb{X})$ from observations of $r_f$ for each method.}
    \label{score_noise}
  \end{subfigure}  

  \caption{Prediction of the three outputs from observations of $r_f$, with a standard deviation of $0.3$ for the measurement noise.}
  \label{fig_smallnoise}

\end{figure}

\Cref{fig_smallnoise} shows the results obtained from measurements of $r_f$ with a noise standard deviation of $0.3$, instead of $0.9$ in the core of the article. The difference between the \textit{Uniform error} approach and the two hierarchical model-based methods is less pronounced, as the prior has a reduced influence.
\end{document}